\pdfoutput=1
\documentclass[10pt,a4paper]{article}

\usepackage[T1]{fontenc}
\usepackage[utf8]{inputenc}
\usepackage{lmodern}
\usepackage[a4paper,margin=2.2cm]{geometry}
\usepackage{amsmath,amssymb}
\usepackage{graphicx}
\usepackage{booktabs,multirow,array}
\usepackage{xcolor}
\usepackage[authoryear,round]{natbib}
\usepackage{longtable}
\usepackage{tabularx}
\usepackage{xurl}
\usepackage{authblk}
\usepackage[colorlinks,allcolors=blue]{hyperref}
\usepackage{paperfmt}
\newcommand{\bestvalue}[1]{{\bfseries\boldmath #1}}

\title{A Quantum Support Vector Regression Framework for Missing Data Imputation in Ship Performance Data}

\author[a]{Daniel Maroto Sanchez}
\author[b,$\ast\ast$]{Christian Velasco-Gallego}
\author[a]{Miriam Bastante Chichon}
\author[b]{Jaime Álvarez-Benayas}
\author[c]{Tsoulakos Nikolaos}
\author[a,$\ast$]{J.A. Bravo-Montes}

\affil[a]{HPC \& Quantum, Bull Spain, Madrid, 28040, Spain}
\affil[b]{ARIES Research Group, Nebrija University, Madrid, 28015, Spain}
\affil[c]{Innovation \& Technology Department, Laskaridis Shipping Co. Ltd, Athens, 14562, Greece}
\date{}

\begin{document}

\maketitle

{\let\thefootnote\relax
\footnotetext{$^{\ast}$Corresponding author}
\footnotetext{$^{\ast\ast}$Principal corresponding author}
\footnotetext{Email addresses: \href{mailto:cvelasco@nebrija.es}{cvelasco@nebrija.es} (C. Velasco-Gallego); \href{mailto:andres.bravo@bull.com}{andres.bravo@bull.com} (J.A. Bravo-Montes)}}

\begin{abstract}
The Internet of Ships enables vessels to generate vast amounts of data, yet missing values remain a persistent issue. To the best of the authors' knowledge, existing imputation techniques in this domain rely exclusively on classical machine learning. This study investigates quantum support vector regression (QSVR) for imputing missing values in ship performance data. Classical support vector regression and QSVR models based on different quantum feature maps are compared for estimating marine fuel consumption under varying levels of data availability. The case study uses real-world data from a Panamax bulk carrier, covering 37 variables across six categories. Three experiments are conducted: exploratory analysis with three input features, scalability analysis with increasing dimensionality (3 to 18 features), and comparative evaluation under a selected representative configuration. Across all scenarios, both classical and quantum approaches captured the relationship between the inputs and the main-engine fuel-flow rate, maintaining strong imputation performance even under severe data scarcity. \textit{ZZFeatureMap\_1} achieved the highest average \(R^2\) (0.991 in the performance comparison conducted with 60\% of the dataset, 90\% missing data, and 14 input features, for instance) and the lowest aggregate imputation errors, while the radial basis function kernel was the strongest classical baseline, delivering highly competitive results.

\end{abstract}

\section{Introduction}
\label{sec:introduction}

In recent years, the maritime sector has undergone a significant transformation driven by advances in artificial intelligence \citep{WANG2024100123}. Vessels have become increasingly interconnected through the Internet of Ships, enabling the deployment of infrastructure capable of generating vast amounts of data \citep{Aslam2020}.

These data have been widely employed to improve operations in areas such as route planning \citep{LI2026125165,jmse14010043}, predictive maintenance \citep{info17010016}, structural health monitoring \citep{SHRIVASTAVA2025110632, YAO2025120277}, and accident investigation \citep{CAO2023113563, Li02012023}. However, the industry still faces significant challenges in ensuring data quality \citep{app13179742}. Nevertheless, this issue remains comparatively underexplored.

Among the various data quality issues, this study focuses on missing values, which may arise from network failures, device failures, or human error \citep{10287194}. If left unaddressed or inadequately handled, these gaps can compromise the reliability and accuracy of results, introduce bias into subsequent analyses, and reduce the quality of models used to support decision-making \citep{VELASCOGALLEGO2020108261}.

Although several studies have investigated missing data in the maritime domain, to the best of the authors' knowledge, the use of quantum machine learning (QML), specifically quantum support vector regression (QSVR), for imputing ship performance data remains unexplored. Quantum kernel methods, such as those underlying QSVR, map data into high-dimensional Hilbert spaces, providing alternative representations for modelling complex, nonlinear relationships in sensor data. This motivates their exploration for imputation tasks \citep{Schuld2019, Havlicek2019}.

Accordingly, this study presents the implementation and evaluation of QSVR for imputing missing values under a missing completely at random (MCAR) mechanism. This mechanism was adopted to enable a controlled evaluation of imputation performance without the additional dependencies introduced by missing at random (MAR) or missing not at random (MNAR) mechanisms. This choice is consistent with common practice when validating novel imputation approaches \citep{Albu2025missForestPredict}.

The remainder of this paper is structured as follows. Section~\ref{sec:related-work} presents related work on machine learning for data imputation and QML applications in the maritime sector. Section~\ref{sec:methodology} describes the proposed methodology and the classical support vector regression (SVR) and QSVR approaches. Section~\ref{sec:dataset-description} describes the dataset and case study, while Section~\ref{sec:experimental_environment} details the software and hardware environments. Section~\ref{sec:results} reports the experimental results, and Section~\ref{sec:conclusions} presents the conclusions.

\section{Related work}
\label{sec:related-work}

\subsection{Missing data imputation methods in maritime applications}
\label{subsec:maritime-imputation}

To identify studies on data imputation techniques in the maritime industry, the following search query was used in the Scopus database:

\par\nobreak\smallskip
\noindent\makebox[\columnwidth][c]{\texttt{"data imputation" AND "maritime"}}
\par\smallskip

The search was conducted in September 2026 using the title, abstract, and keywords (TITLE-ABS-KEY) fileds of the Scopus database. The query returned 23 papers, of which 15 were retained after an initial screening. Papers were excluded primarily because they did not address data imputation in sufficient detail. Table~\ref{tab:imputation_studies} summarises the selected studies.

\makeatletter
\let\prism@original@dbflt\@dbflt
\def\@dbflt#1{\@ifnextchar[{\@xdblfloat{#1}}{\edef\@tempa{\noexpand\@xdblfloat{#1}[\csname fps@#1\endcsname]}\@tempa}}
\makeatother
\begin{ptable}[raw=true,float=true,span=double,pos=!b]{Summary of studies related to the application of data imputation methodologies in the maritime sector. CNN: convolutional neural network; kNN: $k$-nearest neighbours; LSTM: long short-term memory.}{tab:imputation_studies}{}
\small
\begin{tabular}{p{4.2cm} p{5cm} p{4.5cm}}
\toprule
\textbf{Reference} & \textbf{Data Imputation Method} & \textbf{Type of Data} \\
\midrule
\citet{LU2026105016} & GAN-based method & Noon reports and AIS data \\
\citet{XUAN2026125563} & Graph-enhanced Transformer & AIS data \\
\citet{CELIK2026103714} & Fourier-based Compressed Sensing (F-CS) model & Buoy data \\
\citet{WANG2025107279} & Conditional diffusion & AIS data \\
\citet{pr13072035} & Random Forest & Ship maintenance data \\
\citet{Muhammed_Nor_2025} & MICEForest, kNN, and MissForest & Environmental Performance Index (EPI) data \\
\citet{CHAKRABORTY2025120145} & LSTM and attentive neural networks (CNN--LSTM) & Buoy data \\
\citet{SUN2025119722} & Stepwise Multiple Nonlinear Regression & Ship static parameters \\
\citet{jmse12010097} & Multiple Imputation by Chained Equations & Ship data \\
\citet{s23104581} & k-Nearest Neighbours & Satellite and local in-situ data (offshore wind farms) \\
\citet{jmse11040806} & Generative Adversarial Networks & Ship data \\
\citet{10_5957_JOSR_09210032} & Variational Autoencoder & Ship data \\
\citet{Velasco-Gallego03082022} & First-order Markov chain and machine learning & Ship data \\
\citet{Wang2021} & Expectation-Maximization with Bootstrap & Rescue data \\
\citet{VELASCOGALLEGO2020108261} & 20 widely implemented machine learning and time series forecasting algorithms & Ship data \\
\bottomrule
\end{tabular}
\end{ptable}
\makeatletter
\let\@dbflt\prism@original@dbflt
\makeatother

As detailed in Table~\ref{tab:imputation_studies}, the method utilised for data imputation varies according to the type of data being imputed. Across the analysed studies, three main data types were identified: Automatic Identification System (AIS) data, buoy data, and ship data.

AIS is a system that enables the tracking of vessels and contains static information, dynamic information, and voyage-related physical information, thus offering a comprehensive depiction of vessel motion patterns and serving as the essential foundation for trajectory analysis and maritime behaviour modelling \citep{XUAN2026125563}. Accordingly, these data are utilised in a wide variety of applications, such as the management of vessels in waterways, the protection of the environment, and overall surveillance to enhance safety \citep{EMMENS2021114975}. However, despite its capability of retrieving data that can be analysed to provide insight for different purposes, several factors, such as equipment malfunctions, network congestion, and adverse sea conditions, give rise to problems in data collection, thus resulting in missing gaps that may compromise the accuracy of maritime analysis. Accordingly, the application of data-driven methods for adequately completing the dataset is of preeminent importance in order to obtain complete, high-quality vessel trajectories. In this sense, several data imputation methods have been proposed. For instance, a generative adversarial network (GAN) has been used to impute missing values in noon reports and AIS records before both sources are fused for fuel consumption prediction \citep{LU2026105016}. A graph-enhanced Transformer framework has also been developed to address long-range gaps in AIS trajectories \citep{XUAN2026125563}. In addition, a cyclic denoising trajectory imputation method based on conditional diffusion models has been introduced \citep{WANG2025107279}. Thus, as can be observed, the data imputation methods implemented to deal with missing gaps in AIS data aim to capture global spatiotemporal patterns in order to enhance the learning of trajectory distributions and motion features.

The second type of data identified in the studies is buoy data, which contribute to modelling and understanding global weather and climate patterns by providing real-time measurements of meteorological and oceanographic parameters, such as wind, waves, air pressure, and sea temperature \citep{REISS202163}. In this instance, a total of two studies were identified. The first relates to the implementation of hybrid models that decompose complex signals into simpler patterns using mathematical transforms. Specifically, a Fourier-based Compressed Sensing model was introduced to identify the underlying periodic trends and seasonal cycles of the ocean, providing a stable foundation for reconstructing the missing data points. The case study was conducted specifically on the significant wave height parameter \citep{CELIK2026103714}. The second study, analogous to the previous one, decomposes the wave surface elevation into slowly varying amplitudes using a wave model. Then, non-linear data-driven methods, as well as linear methods, are utilised for imputing continuous gaps of varying lengths \citep{CHAKRABORTY2025120145}. Consequently, when imputing missing meteorological and oceanographic data, the signal is usually first decomposed to identify underlying trends, which then serve as the input for a data-driven model used to conduct the reconstruction.

The third main data type is ship data, which appears to be the most frequent type in the analysed studies. Based on the analysed case studies, these data mainly relate to performance and marine machinery data, which are commonly used to enable predictive maintenance. As with the previous data types, missing data pose a significant challenge when conducting quantitative research in this sector and may arise from device failure, network collapse, or human error. For instance, a genetic algorithm--ant colony optimisation collaborative random forest has been used to resolve the local optima issue in high-dimensional missing data imputation \citep{pr13072035}. A stepwise nonlinear regression has also been proposed for ship static data imputation \citep{SUN2025119722}. Furthermore, Multiple Imputation by Chained Equations has been employed \citep{jmse12010097}. Thus, when imputing data in this context, classical machine learning methods are most commonly applied. Models that capture temporal patterns are also frequently implemented.

Consequently, based on the related-work analysis and to the best of the authors' knowledge, there is no evidence that QML, specifically QSVR, has been evaluated for the imputation of ship performance data. Addressing this gap is therefore the main contribution of this study.

\subsection{Quantum machine learning in maritime applications}
\label{subsec:qml-maritime}

To identify publications on QML applications in the maritime sector, the following search query was used in the Scopus database:

\begin{center} \texttt{("quantum machine learning" OR "quantum neural network" OR "quantum support vector machine" OR "quantum kernel" OR "hybrid quantum-classical") AND "maritime"}
\end{center}

Analogous to the previous subsection, the search was conducted in September 2026 using the title, abstract, and keywords (TITLE-ABS-KEY) fields of the Scopus database. The query returned 10 records. After title and abstract screening, only 2 studies satisfied the inclusion criteria, namely the explicit implementation of QML techniques for maritime applications. Records corresponding to review papers, quantum communication systems, optimisation approaches without QML components, or studies that merely discussed QML as a future technology were excluded.

One of the identified studies developed a hybrid quantum-classical neural network architecture for vessel classification from multispectral satellite imagery \citep{Barretta2025Vessel}. Their framework combined a classical ResNet feature extractor with quantum layers and demonstrated the feasibility of integrating quantum components into maritime surveillance workflows.

The other identified study investigated quantum-enhanced deep reinforcement learning and quantum neural networks for resource allocation in maritime 6G-enabled satellite-air-ground integrated networks \citep{Prabhashana2026MaritimeQDRL}. They showed that quantum-enhanced agents achieved lower system costs than their classical counterparts in complex optimisation scenarios.

Overall, the results indicate that research on QML in maritime applications remains limited and is primarily focused on classification and optimisation tasks. This observation is consistent with a recent review that identified maritime QML as an emerging field with a small number of application-specific studies \citep{Schetakis2025Review}.

No studies were found that apply QML techniques to data-quality enhancement, missing-data reconstruction, or imputation problems. To the best of the authors' knowledge, no previous study has investigated the use of QSVR for the imputation of ship-performance data, nor were any QML-based maritime imputation studies identified during the literature review. This gap motivates the research presented in this work.

\section{Methodology}
\label{sec:methodology}

{This section presents the methodology adopted to benchmark classical and quantum approaches for marine data imputation. The experimental workflow, illustrated in Figure~\ref{fig:methodology-workflow}, comprises four main stages. First, the data preprocessing procedure is described, including the preparation of the dataset for both classical and quantum models (Section~\ref{subsec:datapreprocessing}). Second, the modelling framework is introduced, covering both the classical SVR baseline and its quantum counterpart, QSVR, which replaces the classical kernel with one estimated through quantum feature maps (Section~\ref{subsec:modelling}). Third, a unified hyperparameter tuning strategy is defined to ensure a controlled and fair comparison between all configurations (Section~\ref{subsec:hyperparameter-tuning}). Finally, the evaluation framework used to assess imputation performance is presented (Section~\ref{subsec:evaluation}).}

\makeatletter
\let\prism@original@dbflt\@dbflt
\def\@dbflt#1{\@ifnextchar[{\@xdblfloat{#1}}{\edef\@tempa{\noexpand\@xdblfloat{#1}[\csname fps@#1\endcsname]}\@tempa}}
\makeatother
\begin{pfig}[span=double,pos=!b]{Overview of the experimental workflow used to compare classical SVR and QSVR models for missing-data imputation.}{fig:methodology-workflow}
    \includegraphics[width=.75\textwidth]{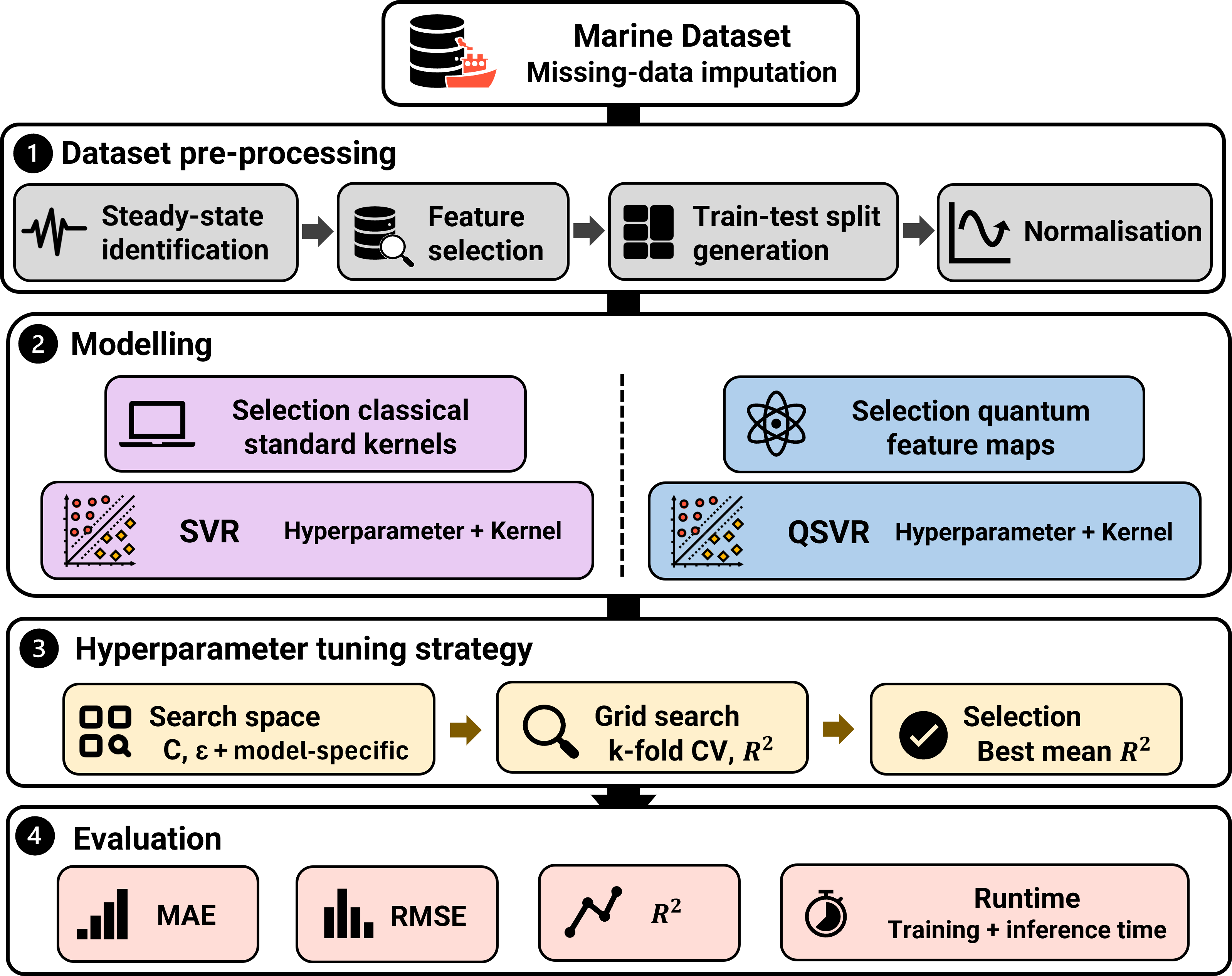}
\end{pfig}
\makeatletter
\let\@dbflt\prism@original@dbflt
\makeatother

\subsection{Dataset preprocessing}
\label{subsec:datapreprocessing}
Data preprocessing has been minimised in order to adequately assess the data imputation capability of the implemented methods when dealing with raw sensor data. However, certain steps must first be carried out to perform this task adequately. These steps are steady-state identification (Section~\ref{subsubsec:steady-states}), feature selection (Section~\ref{subsubsec:feature-selection}), train-test split generation (Section~\ref{subsubsec:train-test-generation}), and normalisation (Section~\ref{subsubsec:normalisation}).

\subsubsection{Identification of steady states}
\label{subsubsec:steady-states}
Raw data typically include non-operational states that negatively affect the results obtained when applying data-driven approaches. Although marine engines generally operate under steady-state conditions, variations can arise, for example, due to environmental factors or changes in the operating mode. Consequently, if these states are not properly handled, both computational efficiency and model performance may decline.

To address this matter, the sliding-window $t$-test algorithm is applied to the main engine rotational speed feature for steady-state detection. Thus, it is assumed that engine performance metrics are quasi-stationary during steady-state operation but exhibit pronounced non-stationarity during transients (acceleration, deceleration, manoeuvring).

To do so, the rpm signal is partitioned into non-overlapping windows of length $m$. For each pair of consecutive windows, the sample means and sample variances are computed. Then, the following statistic is formed:

\begin{equation}
t = \frac{|\bar{x}_1 - \bar{x}_2|}{\sqrt{\dfrac{s_1^2}{m} + \dfrac{s_2^2}{m}}},
\end{equation}

where $\bar{x}_1$ and $\bar{x}_2$ are the sample means, and $s_1^2$ and $s_2^2$ are the sample variances.

The resulting statistic is compared against a two-tailed critical value obtained from Student's $t$-distribution with $2m-2$ degrees of freedom at a significance level of $\alpha=0.05$:

\begin{equation}
t_{crit} = t_{0.975,\ 2m-2}
\end{equation}

The null hypothesis $H_0: \mu_1 = \mu_2$ is retained when $t < t_{crit}$, indicating no statistically significant difference between the two adjacent windows' means, and thus, that both samples pertain to the same operational steady state. Conversely, when the null hypothesis is rejected, it is implied that there is a significant shift in the local mean, and it is thus considered that a change in the operational steady state has occurred.

To avoid steady states at low rpms, which may indicate that the engine is off or in a low-load transitional state, a validity threshold is also set to discard instances below such a threshold. Additionally, windows containing missing values are directly flagged as non-steady. Finally, candidate steady segments shorter than a minimum run length are discarded and reclassified as non-steady, removing spurious short segments that arise near transients.

\subsubsection{Feature selection}
\label{subsubsec:feature-selection}
Selecting an adequate set of features is of paramount importance in this task, since each variable is mapped to a qubit. Increasing the number of variables therefore increases the computational cost of the model, and beyond a certain point, the number of qubits required may exceed the available computational resources, making the model unfeasible to execute. This selection has been made based on domain knowledge in an attempt to predict from independent causal inputs to avert target leakage.

\subsubsection{Train-test split generation}
\label{subsubsec:train-test-generation}
The resulting dataset from the previous step is split into training and test sets to adequately evaluate the data imputation performance of the implemented methods. Additionally, cross-validation is also conducted to obtain a more reliable estimate of the evaluated performance.

As with feature selection, the number of instances in the dataset may constrain the practical feasibility of implementing the QML model being analysed. Accordingly, an additional study is conducted to determine the amount of data required to train the model adequately. To this end, the dataset is randomly partitioned into subsets of varying size, ranging from 10\% to 100\% of the total dataset in increments of 10\%. This procedure yields ten subsets, corresponding to 10\%, 20\%, and so on up to 100\% of the original data. This analysis enables the minimum viable amount of training data to be identified.

\subsubsection{Normalisation}
\label{subsubsec:normalisation}
As the SVR model relies on kernel functions, such as the Radial Basis Function (RBF), which measure distance or similarity between instances, feature scaling is required. Accordingly, normalisation is conducted by applying the following equation:

\begin{equation}
x' = \frac{x - x_{min}}{x_{max} - x_{min}}
\end{equation}

The resulting features lie in the $[0,1]$ range. The same normalisation is applied to the classical and quantum models to ensure a consistent comparison; it is additionally required by the quantum feature maps, whose angle-based encodings map feature values onto bounded rotation angles.

\subsection{Modelling}
\label{subsec:modelling}
{Both classical and quantum models are formulated within the SVR framework, which provides a natural and controlled common backbone for this comparison, as illustrated in Figure~\ref{fig:svr-framework}. SVR suits trajectory imputation well: the $\epsilon$-insensitive loss confers robustness to outliers, the kernel trick captures non-linear structure, and the method generalises reliably in small-to-medium sample regimes \citep{Vapnik1995, Smola2004, Drucker1997}. The classical SVR model serves as the baseline, using standard kernels defined in closed form. The quantum extension, QSVR, replaces the classical kernel with one estimated via quantum feature maps, allowing the model to exploit high-dimensional Hilbert spaces as the implicit feature space. Sections~\ref{subsubsec:svr} and~\ref{subsubsec:qsvr} describe the classical and quantum models, respectively.}

\begin{pfig}[span=double,pos=t]{Common modelling framework for classical SVR and quantum-kernel-based QSVR.}{fig:svr-framework}
    \includegraphics[width=\textwidth]{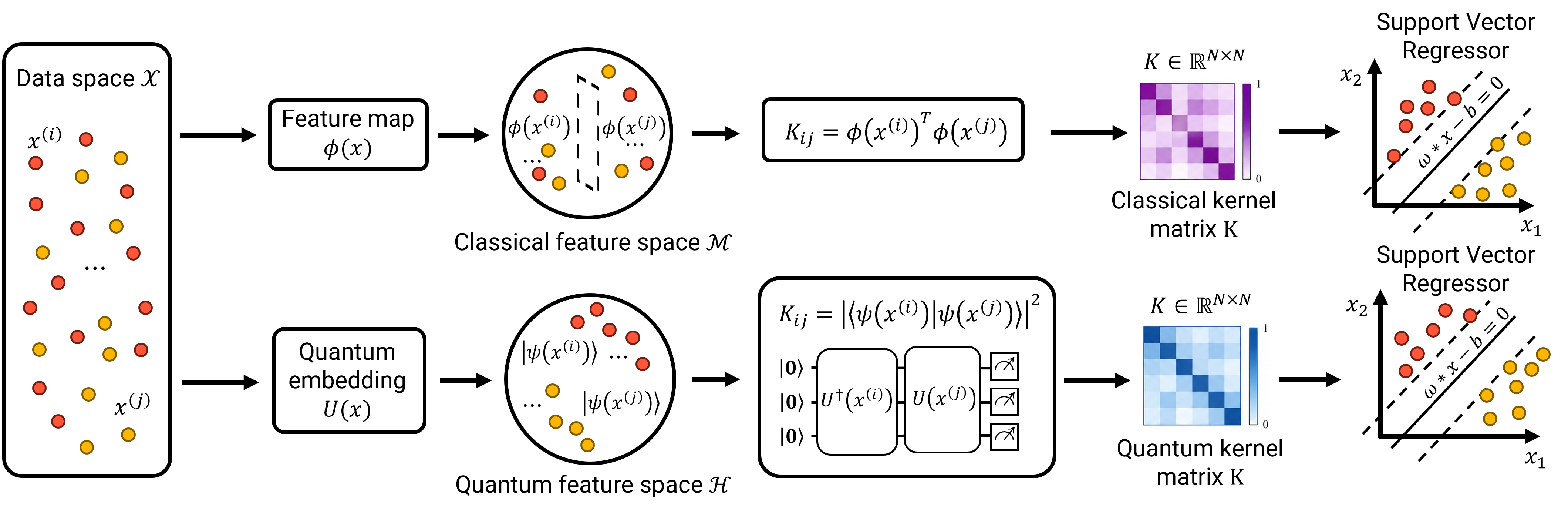}
\end{pfig}

\subsubsection{Support Vector Regression}
\label{subsubsec:svr}
{
The classical SVR model constitutes the baseline of this study and provides the common backbone that is subsequently extended to its quantum counterpart (Section~\ref{subsubsec:qsvr}). SVR adapts the margin-based principles of support vector machines to real-valued function estimation \citep{Vapnik1995, Drucker1997}. It retains their defining properties: a sparse solution expressed through a subset of the training samples, control of model complexity through margin maximisation, and the capacity to represent non-linear relationships by means of the kernel trick \citep{Smola2004, Awad2015}.

Given a training set $\{(x_i, y_i)\}_{i=1}^{N}$, with input vectors $x_i = (x_i^{(1)}, \dots, x_i^{(d)}) \in \mathbb{R}^{d}$ and targets $y_i \in \mathbb{R}$, $N$ represents the number of samples, $d$ the number of features, the subscript $i$ identifies the data point and the superscript identifies the feature. SVR seeks a regression function of the form
\begin{equation}
\label{eq:svr-function}
f(x) = \langle w, \phi(x) \rangle + b,
\end{equation}
where $\phi(\cdot)$ maps the inputs into a (possibly high-dimensional) feature space, $w$ is the weight vector, and $b$ is the bias term.

SVR employs Vapnik's $\epsilon$-insensitive loss function, which ignores residuals smaller than a tolerance $\epsilon$ and grows linearly beyond it \citep{Vapnik1995}:
\begin{equation}
\label{eq:epsilon-loss}
L_{\epsilon}\big(y, f(x)\big) = \max\big(0,\; |y - f(x)| - \epsilon \big).
\end{equation}
This defines a tube of radius $\epsilon$ around the estimated function, within which the residuals incur no penalty, endowing the model with robustness to small perturbations and outliers.

The model parameters are obtained by solving the convex optimisation problem
\begin{equation}
\label{eq:svr-primal-obj}
\min\limits_{\textstyle w,\,b,\,\xi,\,\xi^{*}} \quad \frac{1}{2}\|w\|^{2} + C\sum_{i=1}^{N}\big(\xi_i + \xi_i^{*}\big),
\end{equation}
subject to
\begin{equation}
\label{eq:svr-primal-constraints}
\begin{aligned}
y_i - \langle w, \phi(x_i) \rangle - b &\leq \epsilon + \xi_i, \\
\langle w, \phi(x_i) \rangle + b - y_i &\leq \epsilon + \xi_i^{*}, \\
\xi_i,\; \xi_i^{*} &\geq 0, \qquad i = 1,\dots,N,
\end{aligned}
\end{equation}
where $\xi_i$ and $\xi_i^{*}$ are slack variables measuring deviations above and below the $\epsilon$-tube, respectively, and the regularisation constant $C > 0$ governs the trade-off between the flatness of $f$ (i.e. the norm $\|w\|$) and the tolerance to deviations larger than $\epsilon$.

Introducing Lagrange multipliers $\alpha_i, \alpha_i^{*}$ and applying the Karush--Kuhn--Tucker conditions yields the dual formulation \citep{Smola2004}:
\begin{equation}
\label{eq:svr-dual}
\begin{aligned}
\max\limits_{\textstyle \alpha,\,\alpha^{*}} \quad & -\frac{1}{2}\sum_{i,j=1}^{N}(\alpha_i - \alpha_i^{*})(\alpha_j - \alpha_j^{*})\, K(x_i, x_j) \\
& -\,\epsilon\sum_{i=1}^{N}(\alpha_i + \alpha_i^{*}) + \sum_{i=1}^{N} y_i(\alpha_i - \alpha_i^{*}),
\end{aligned}
\end{equation}
subject to $\sum_{i=1}^{N}(\alpha_i - \alpha_i^{*}) = 0$ and $0 \leq \alpha_i, \alpha_i^{*} \leq C$. The mapped samples enter the problem only through the inner product $K(x_i, x_j) = \langle \phi(x_i), \phi(x_j) \rangle$, so the feature map $\phi$ never needs to be evaluated explicitly; which is known as the kernel trick.

Once the dual problem is solved, the prediction for a new input $x$ is
\begin{equation}
\label{eq:svr-prediction}
f(x) = \sum_{i=1}^{N}(\alpha_i - \alpha_i^{*})\, K(x_i, x) + b.
\end{equation}
Only the samples with non-zero coefficients $(\alpha_i - \alpha_i^{*})$, known as support vectors, contribute to the prediction.

The choice of kernel $K$ determines the geometry of the implicit feature space and is therefore the principal modelling decision. For the classical baseline, the standard closed-form kernels (linear, RBF, polynomial, and sigmoid) are considered; they are detailed together with their search spaces in Section~\ref{subsec:hyperparameter-tuning} (Table~\ref{tab:classic_hyperparameters}). This same formulation is the point of departure for the quantum model: the QSVR retains the dual problem in Eq.~\eqref{eq:svr-dual} and the prediction rule in Eq.~\eqref{eq:svr-prediction} unchanged, and replaces only the kernel function $K$ with one estimated from quantum feature states (Section~\ref{subsubsec:qsvr}). In this way, the classical and quantum models differ exclusively in how the similarity between samples is computed.
}

\subsubsection{Quantum Support Vector Regression }
\label{subsubsec:qsvr}
{
As mentioned before, designing an effective kernel for complex datasets can be challenging. QSVR models aim to address this limitation by exploiting quantum-mechanical feature spaces, in which classical data are encoded into quantum states and pairwise similarities are evaluated via quantum circuits \citep{Havlicek2019}.
The central idea behind a QSVR is the use of a quantum feature map \citep{Schuld2019}. The quantum feature map, denoted by $\Phi$, transforms a classical data point $x_i$ into a quantum state $|\Phi(x_i)\rangle$. Throughout this paper, $d$ denotes the number of input features and $n$ the number of qubits. Each feature is encoded on one qubit, so $n=d$. Given an encoding circuit $U_{\Phi}(x_i)$ acting on an initial state of $n$ qubits, the resulting quantum feature state is

\begin{equation}
|\Phi(x_i)\rangle = U_{\Phi}(x_i)|0\rangle^{\otimes n}.
\end{equation}

The similarity between two samples $x_i$ and $x_j$ is then quantified through a quantum kernel defined as the overlap between their corresponding quantum states,

\begin{equation}
K(x_i,x_j) = |\langle \Phi(x_i) | \Phi(x_j) \rangle|^2.
\end{equation}

By evaluating the quantum kernel for all pairs of training samples, a kernel matrix is constructed and subsequently used within a classical SVR optimisation procedure. In this hybrid formulation, the quantum component is restricted to the estimation of kernel entries via quantum feature states, whereas the optimisation and prediction stages are performed classically.

Several quantum feature maps have been proposed in the literature, differing in the way classical information is encoded and how correlations between features are introduced. While established quantum feature maps such as ZZFeatureMap, Angle Encoding, and Hardware-Efficient Ansatzes remain widely used as benchmark encodings, recent research has increasingly focused on developing task-specific and data-aware quantum kernels tailored to particular data structures and learning objectives \citep{Glick2024,Matsumoto2026}.

In this work, however, the objective is to compare quantum models against classical baselines and to explore the use of QML for data imputation in ship performance data. Standard Angle Encoding produces a separable kernel with no cross-feature correlations; it is therefore augmented with a configurable entangling layer. For the ZZFeatureMap, the standard interaction function is used. Sections~\ref{subsubsec:angle-encoding} and~\ref{subsubsec:zzfeaturemap} describe these feature maps and their implementation in this work.

One key component of quantum circuits is the entanglement between qubits, which determines the correlation between features and can be introduced by gates acting on two or more qubits. Both feature maps use the same set of configurable entanglement topologies. The topology only determines the qubit pairs over which two-qubit interactions are applied, as summarised in Table~\ref{tab:entanglement-topologies}; the gate applied to each pair is feature-map specific.

\begin{ptable}[raw=true,float=false]{Shared entanglement topologies used by Angle Encoding and ZZFeatureMap.}{tab:entanglement-topologies}{}
\small
\par\vspace{2pt}
\begin{tabular}{ll}
\hline
\textbf{Topology} & \textbf{Pair set $\mathcal{E}_T$} \\
\hline
Linear & $\{(k,k+1): k=1,\dots,n-1\}$ \\
Circular & $\{(k,k+1): k=1,\dots,n-1\} \cup \{(n,1)\}$ \\
Full & $\{(k,l): 1 \leq k < l \leq n\}$ \\
\hline
\end{tabular}
\end{ptable}

For a selected entanglement topology $T$ and two-qubit gate family $G$, the topology-dependent interaction layer can be written generically as
\begin{equation}
\label{eq:shared-entanglement-layer}
U_{\text{ent}}^{(G,T)}(x_i) = \prod_{(k,l) \in \mathcal{E}_T} G_{k,l}(x_i),
\end{equation}
where $G_{k,l}(x_i)=CZ_{k,l}$ for Angle Encoding and $G_{k,l}(x_i)=R_{ZZ}^{(k,l)}(x_i)$ for ZZFeatureMap.

\subsubsection{Angle Encoding}
\label{subsubsec:angle-encoding}
Angle Encoding is a feature map that embeds a classical data point
$x_i \in \mathbb{R}^d$ into an $n$-qubit quantum state using single-qubit rotations.
Entanglement may be introduced subsequently as an additional layer in composite
feature map constructions.

In this work, this extended construction is adopted by adding an entangling layer
based on controlled-Z gates in order to capture correlations between
different features and enrich the resulting quantum representation.

Given a data point $x_i = (x_i^{(1)}, \dots, x_i^{(d)})$, the feature map defines a unitary transformation $U_{\Phi}(x_i)$ acting on the initial state $|0\rangle^{\otimes n}$:

\begin{equation}
\label{eq:angle-feature-state-general}
|\Phi(x_i)\rangle = U_{\Phi}(x_i) |0\rangle^{\otimes n}.
\end{equation}

The implemented feature map consists of three consecutive layers. A first rotation layer encodes each feature $x_i^{(k)}$ onto its corresponding qubit through a Pauli rotation gate $R_{\alpha}(x_i^{(k)})$, where $\alpha \in \{x,y,z\}$ selects the rotation axis:
\begin{equation}
\label{eq:angle-rotation-layer}
U_{\text{rot}}^{\alpha}(x_i) = \prod_{k=1}^{d} R_{\alpha}(x_i^{(k)}), \quad \alpha \in \{x,y,z\}.
\end{equation}

An entangling layer then introduces correlations between qubits using controlled-$Z$ interactions over the shared topology $T$ defined in Table~\ref{tab:entanglement-topologies}. Since the entangling layer would otherwise leave the kernel unchanged, a second rotation layer is appended, resulting in the circuit shown in Figure~\ref{fig:angleencoding} with its complete Angle Encoding unitary given by:
\begin{equation}
\label{eq:angle-feature-map-unitary}
U_{\Phi}^{(\alpha,T)}(x_i) = U_{\text{rot}}^{\alpha}(x_i)U_{\text{ent}}^{(CZ,T)}(x_i) \, U_{\text{rot}}^{\alpha}(x_i).
\end{equation}

\begin{pfig}[float=false]{Representation of the implemented Angle Encoding feature map with a configurable entanglement layer.}{fig:angleencoding}
\centering
\includegraphics[width=0.8\columnwidth]{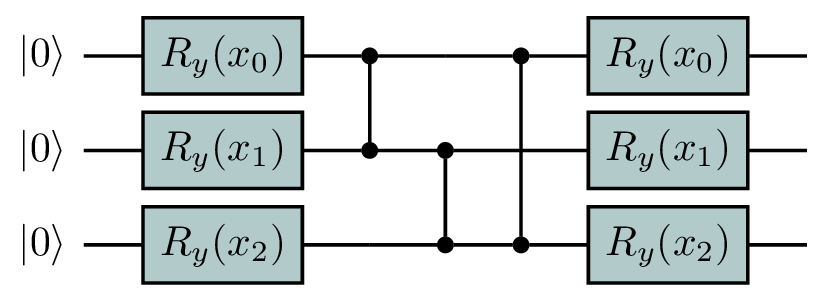}
\end{pfig}

\subsubsection{ZZFeatureMap}
\label{subsubsec:zzfeaturemap}

The ZZFeatureMap encodes classical data into a quantum state through a combination
of single-qubit phase rotations and data-dependent pairwise interactions.
Compared to rotation-based feature maps, it introduces explicit non-linear feature
couplings through a ZZ-type interaction structure, leading to richer entanglement
patterns and higher expressivity.

The implemented ZZFeatureMap consists of repeated blocks, referred to as repetitions. Each repetition comprises two layers. In the single-qubit encoding layer, each qubit is first prepared in superposition by a Hadamard gate and then subjected to a data-dependent phase rotation:
\begin{equation}
\label{eq:zz-encoding-layer}
U_{\text{enc}}(x_i) =
\prod_{k=1}^{d} R_Z(2x_i^{(k)}) \, H_k.
\end{equation}
This step embeds each feature into the phase of the quantum state.

The ZZ entangling layer then introduces correlations through pairwise $ZZ$ rotations, implemented via the controlled-NOT (CNOT)--$R_Z$--CNOT decomposition. For each pair $(k,l)$, a data-dependent interaction is applied with rotation angle set by the interaction function $\phi(x_i^{(k)}, x_i^{(l)})$:
\begin{equation}
\label{eq:zz-interaction}
R_{ZZ}^{(k,l)} =
\exp\big(-i \, \phi\big(x_i^{(k)}, x_i^{(l)}\big) \, Z_k Z_l \big).
\end{equation}
Each feature is encoded on its own qubit, so feature index and qubit
index coincide; $Z_k$ denotes the Pauli-$Z$ operator on the qubit
encoding feature $k$.

In this work, the ZZFeatureMap uses the standard interaction function \citep{Havlicek2019}:
\begin{equation}
\label{eq:zz-standard-interaction}
\phi(x_i^{(k)}, x_i^{(l)}) = (\pi - x_i^{(k)})(\pi - x_i^{(l)}).
\end{equation}
This function determines the data-dependent coupling between each pair of interacting qubits and is fixed throughout the experiments.

The structure of the ZZFeatureMap is further determined by two quantum experiment configuration choices: the shared entanglement topology (same decision as in Table \ref{tab:entanglement-topologies}) and the number of repetitions.

Repeating the single-qubit encoding and ZZ entangling layers $L$ times results in the implementation schema shown in Figure~\ref{fig:zzfeaturemap} and gives the complete ZZFeatureMap unitary:
\begin{equation}
\label{eq:zz-unitary-fixed}
U_{\Phi}^{(\phi,T,L)}(x_i) =
\left[
U_{\text{ent}}^{(R_{ZZ},T)}(x_i) \, U_{\text{enc}}(x_i)
\right]^L.
\end{equation}
}

\begin{pfig}[float=false]{Representation of the implemented ZZFeatureMap for a linear
entanglement pattern.}{fig:zzfeaturemap}
\centering
\includegraphics[width=\columnwidth]{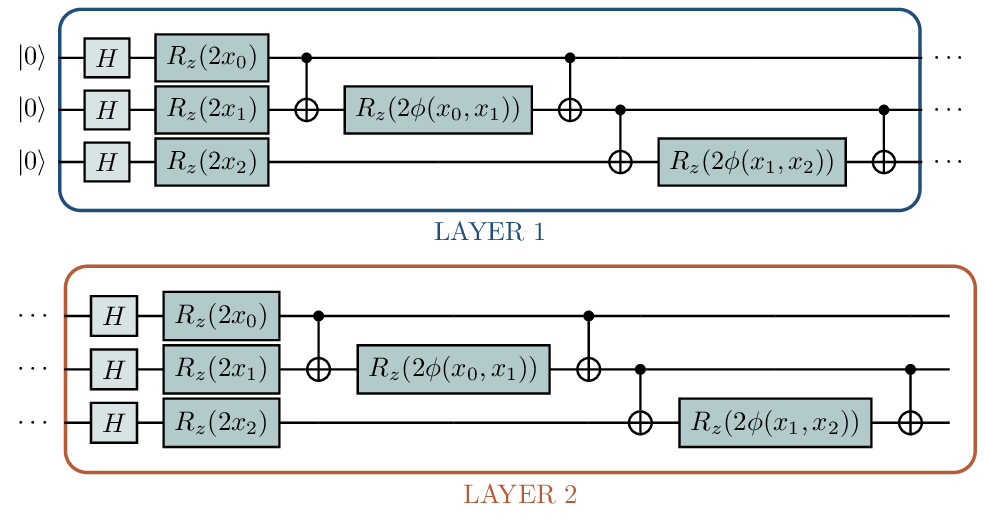}
\end{pfig}

\subsection{Hyperparameter tuning strategy}
\label{subsec:hyperparameter-tuning}
{
All models, both classical and quantum, were built upon the SVR implementation
from \texttt{scikit-learn} \citep{scikit-learn}. Classical kernels are computed in closed form,
whereas quantum kernels are evaluated from quantum feature states
(Section~\ref{subsubsec:qsvr}) and supplied to the solver as a precomputed
matrix (\texttt{kernel='precomputed'}). All remaining settings and the
selection procedure are identical across both families, so that only the
kernel varies. Hyperparameter optimisation was performed using a grid search
over the regularisation parameter $C$ and the $\epsilon$ parameter that defines
the width of the $\epsilon$-insensitive tube. The search space is summarised in Table~\ref{tab:svr-grid-search}, yielding 35 combinations.

\begin{ptable}[raw=true,float=false]{Grid-search space used for SVR hyperparameter tuning. The seven $C$ values are logarithmically spaced.}{tab:svr-grid-search}{}
\small
\par\vspace{1pt}
\begin{tabular}{@{}ll@{}}
\hline
\textbf{Parameter} & \textbf{Search values} \\
\hline
$C$ & $\{10^{4/7}, 10^{8/7}, 10^{12/7}, 10^{16/7}, 10^{20/7}, 10^{24/7}, 10^{4}\}$ \\
$\epsilon$ & $\{1, 5, 10, 25, 50\}$ \\
\hline
\end{tabular}
\end{ptable}

Each configuration was evaluated using 5-fold
cross-validation with shuffled splits (\texttt{random\_state}=42), and the
coefficient of determination ($R^2$) was used as the model selection criterion.
The best-performing configuration was subsequently retrained and evaluated on
the held-out set. Classical-specific kernels and corresponding hyperparameters are detailed in
Section~\ref{subsubsec:classical-hyperparameters}, while the quantum experiment
configurations are presented in Section~\ref{subsubsec:quantum-configurations}.
}

\subsubsection{Classical-specific hyperparameters}
\label{subsubsec:classical-hyperparameters}

Apart from the $C$ and $\epsilon$ parameters, one of the main hyperparameters to tune when implementing the classical SVR is the kernel type. For this study, four types of kernels have been analysed: linear, RBF, polynomial, and sigmoid.

The linear kernel is considered effective when data are linearly separable and can be represented by the following equation \citep{frai_2024_1287875}:

\begin{equation}
K(x_i, x_j) = x_i^T x_j,
\end{equation}

where $x_i$ and $x_j$ are the feature vectors.

The RBF kernel is utilised when data are not linearly separable and aims to measure the degree of similarity between two instances. To do so, the Euclidean distance is considered. This type of kernel can be mathematically represented as follows \citep{DING202110121}:

\begin{equation}
K(x_i, x_j) = \exp\left(-\gamma \|x_i - x_j\|^2\right),
\end{equation}

where $\gamma$ equals $\dfrac{1}{2\sigma^2}$ and determines the radius of influence of each training sample. Because $\gamma$ significantly affects kernel performance, it must be selected appropriately \citep{almejibli2020effect}.

Analogous to the RBF kernel, the polynomial kernel introduces non-linearity into the feature space and can be represented by the following equation:

\begin{equation}
K(x_i, x_j) = \left(\gamma x_i^T x_j + r\right)^p,
\end{equation}

where $\gamma$ controls the influence of the linear part of the kernel, $r$ is the intercept term, and $p$ is the degree of the polynomial \citep{frai_2024_1287875}. All three parameters also need to be tuned when selecting this kernel, and their adequate selection may affect its performance.

The last kernel analysed in this study is the sigmoid kernel. This kernel can be mathematically represented as:

\begin{equation}
K(x_i, x_j) = \tanh\left(\gamma x_i^T x_j + r\right)
\end{equation}

The $\gamma$ and $r$ parameters are also present in this kernel and must be selected appropriately.

Table~\ref{tab:classic_hyperparameters} summarises the search space of each of the hyperparameters considered for implementing the classical SVR. Please note that these search spaces have been selected based on heuristic evaluation and may need to be adapted based on the characteristics of the data. Let $X=[x_1^T,\dots,x_N^T]^T\in\mathbb{R}^{N\times d}$ denote the training design matrix, with $N$ samples and $d$ input features. In Table~\ref{tab:classic_hyperparameters}, $\operatorname{Var}(X)$ denotes the empirical variance of all entries of $X$,

\begin{equation}
\begin{aligned}
\operatorname{Var}(X)
&=\frac{1}{Nd}\sum_{i=1}^{N}\sum_{k=1}^{d}
\left(x_i^{(k)}-\bar{X}\right)^2, \\
\bar{X}
&=\frac{1}{Nd}\sum_{i=1}^{N}\sum_{k=1}^{d}x_i^{(k)}.
\end{aligned}
\end{equation}

\begin{ptable}[raw=true,float=false]{Kernel hyperparameters and search space}{tab:classic_hyperparameters}{}
\begin{adjustbox}{max width=\columnwidth}
\renewcommand{\arraystretch}{1.25}
\begin{tabular}{ll}
\toprule
\textbf{Kernel Type} & \textbf{Hyperparameter and search space} \\
\midrule
Linear     & -- \\
\midrule
RBF        & $\gamma \in [10^{-4}, 10^{0}]$ \\
\midrule
Polynomial & $\gamma \in \left\{\dfrac{1}{d \cdot \text{Var}(X)}, \dfrac{1}{d}\right\}$ \\
           & $p \in \{2, 3\}$ \\
           & $r \in \{0, 1\}$ \\
\midrule
Sigmoid    & $\gamma \in \left\{\dfrac{1}{d \cdot \text{Var}(X)}, \dfrac{1}{d}\right\} \cup [10^{-3}, 10^{-1}]$ \\
           & $r \in \{0, 1\}$ \\
\bottomrule
\end{tabular}
\end{adjustbox}
\end{ptable}

{
\subsubsection{Quantum experiment configurations}
\label{subsubsec:quantum-configurations}
The quantum experiments consider two feature-map families: Angle Encoding and ZZFeatureMap.
The quantum-specific experiment configuration choices are summarised in Table~\ref{tab:quantum-configurations}. These choices define the induced quantum kernels that will be used in the fitting step.

\begin{ptable}[raw=true,float=true, pos=htbp]{Quantum experiment configurations selected for the QSVR models.}{tab:quantum-configurations}{}
\begin{adjustbox}{max width=\columnwidth}
\renewcommand{\arraystretch}{1.2}
\begin{tabular}{lcccc}
\toprule
\textbf{Feature map} & \textbf{Rotation} & \textbf{Topology} & \textbf{Interaction function} & \textbf{Repetitions} \\
\midrule
Angle Encoding & $R_y$ & Linear & --            & --    \\
Angle Encoding & $R_y$ & Full   & --            & --    \\
ZZFeatureMap   & --    & Linear & Standard $\phi$ & $L=1$ \\
ZZFeatureMap   & --    & Linear & Standard $\phi$ & $L=2$ \\
\bottomrule
\end{tabular}
\end{adjustbox}
\end{ptable}

For Angle Encoding, each feature is encoded with an $R_y$ rotation,
\begin{equation}
\label{eq:ry-state}
R_y(x_i^{(k)})\,|0\rangle = \cos(x_i^{(k)}/2)\,|0\rangle + \sin(x_i^{(k)}/2)\,|1\rangle .
\end{equation}
Of the available axes $\alpha \in \{x,y,z\}$, $R_z$ acts on $|0\rangle$ only as a
global phase and therefore encodes no information, leaving every kernel entry
unchanged. Both $R_x$ and $R_y$ can encode information; $R_y$ is adopted because it
produces real-valued amplitudes, matching the real-valued inputs of the data.
Without an entangling layer the feature map factorises over qubits, and the
kernel between two points $x_i$ and $x_j$ reduces to a product of
single-qubit overlaps,
\begin{equation}
\label{eq:angle-factorized-kernel}
K(x_i, x_j) = \prod_{k=1}^{d} \cos^2\!\Big(\tfrac{x_i^{(k)} - x_j^{(k)}}{2}\Big),
\end{equation}
where $x_j^{(k)}$ denotes the $k$-th feature of the second data point $x_j$.
This product carries no cross-feature information, so the entangling layer is the
only mechanism that introduces feature correlations into the kernel. The two extremes
of entanglement topology, linear ($n-1$ gates) and full ($\frac{n(n-1)}{2}$ gates), are
therefore evaluated, leaving the circular topology for future exploration.

For the ZZFeatureMap, the standard interaction function defined in Equation~\ref{eq:zz-standard-interaction} is used with a fixed linear entanglement topology as the simpler default. Finally, $L \in \{1,2\}$ repetitions are used: a larger
$L$ increases expressivity but pushes the off-diagonal kernel values towards a
common constant, reducing the discriminative structure of the kernel
matrix \citep{Thanasilp2024}. The $L=1$ case also probes the more classically
tractable regime, giving a reference for assessing whether the deeper map offers
a real advantage.

}
\subsection{Evaluation}
\label{subsec:evaluation}
{
The classical and quantum models are evaluated using a common set of metrics covering reconstruction quality, generalisation, and computational efficiency: Mean Absolute Error (MAE), Root Mean Squared Error (RMSE), $R^2$, relative overfitting, and runtime. MAE, RMSE, and $R^2$ are computed on the held-out set of injected missing values, comparing the observed ground-truth values $y_i$ with the corresponding model predictions $\hat{y}_i$ over the $N$ evaluation samples. Relative overfitting compares training and test performance, while runtime quantifies the combined computational cost of training and inference.

The same set of performance metrics is applied to every model to ensure a consistent comparison. Reporting complementary measures of reconstruction quality is common practice in maritime data-imputation studies \citep{VELASCOGALLEGO2020108261}. Each measure captures a different aspect of predictive performance, and no single measure is sufficient on its own \citep{Chicco2021, ChaiDraxler2014}. Each performance metric is defined in the following subsections.

\subsubsection{MAE}
\label{subsubsec:metric-mae}
The MAE is the average of the absolute residuals,
\begin{equation}
\label{eq:mae}
\text{MAE} = \frac{1}{N}\sum_{i=1}^{N}\big|\, y_i - \hat{y}_i \,\big|.
\end{equation}
It is expressed in the same units as the target variable and weights all errors linearly, which makes it comparatively robust to large outliers \citep{Willmott2005}.

\subsubsection{RMSE}
\label{subsubsec:metric-rmse}
The RMSE is the square root of the mean squared residual,
\begin{equation}
\label{eq:rmse}
\text{RMSE} = \sqrt{\frac{1}{N}\sum_{i=1}^{N}\big(y_i - \hat{y}_i\big)^{2}}.
\end{equation}
It restores the units of the target variable while retaining the emphasis on large errors, and is regarded as an appropriate summary when the error distribution is expected to be approximately Gaussian \citep{ChaiDraxler2014}.

\subsubsection{Coefficient of Determination ($R^{2}$)}
\label{subsubsec:metric-r2}
The $R^{2}$ score measures the proportion of the target variance explained by the model,
\begin{equation}
\label{eq:r2}
R^{2} = 1 - \frac{\sum_{i=1}^{N}(y_i - \hat{y}_i)^{2}}{\sum_{i=1}^{N}(y_i - \bar{y})^{2}},
\end{equation}
where $\bar{y}$ denotes the mean of the observed values. Being normalised by the variability of the data, $R^{2}$ is a scale-independent goodness-of-fit indicator bounded above by $1$, and has been argued to be more informative than error-only metrics for regression evaluation \citep{Chicco2021}.

\subsubsection{Relative overfitting}
\label{subsubsec:metric-relative-overfitting}
Relative overfitting is measured as the relative difference between the training and test $R^2$ scores:
\begin{equation}
\label{eq:relative-overfitting}
\mathrm{RO} = \frac{R^2_{\mathrm{train}}-R^2_{\mathrm{test}}}{R^2_{\mathrm{train}}},
\end{equation}
where $R^2_{\mathrm{train}}$ is computed on the observed training instances and $R^2_{\mathrm{test}}$ on the masked instances. This expression has the same form as the relative overfitting rate of \citet{EfronTibshirani1997}, $(\theta_{\mathrm{test}}-\theta_{\mathrm{train}})/(\gamma-\theta_{\mathrm{train}})$, with $\theta = R^2$ and the no-information performance set to $\gamma = 0$, the $R^2$ of a predictor that always returns the mean of the target. Values close to zero indicate a small generalisation gap relative to the training fit. Unlike the original definition, $\mathrm{RO}$ is not truncated to $[0,1]$, so negative values indicate a higher test score.

\subsubsection{Runtime}
\label{subsubsec:metric-runtime}
Runtime is the total time, expressed in seconds, required for the training and inference steps for each model. Training time  covers hyperparameter search and model fitting, whereas inference time covers the generation of predictions on unseen inputs. The runtime metric reported in the results is the sum of these two components.
}

\section{Dataset Description}
\label{sec:dataset-description}
The dataset used in this study consists of high-frequency operational telemetry collected from a Panamax bulk carrier operating in international trade over the period from 1 January 2024 to 31 January 2025. The measurements were continuously acquired through the vessel's onboard digital monitoring infrastructure at a temporal resolution of one minute, providing a detailed record of propulsion, machinery, navigation, loading condition, and environmental variables under real operating conditions. Unlike simulation-based datasets frequently employed in maritime digitalisation research, the present dataset originates from an operational merchant vessel and captures actual variations in machinery performance, navigation, loading condition, and the surrounding operating environment.

The study period spans 397 operational days, corresponding to approximately 571,680 observations for each recorded variable prior to data quality control and preprocessing. Table~\ref{tab:vessel-particulars} summarizes the principal particulars of the monitored vessel.

\begin{ptable}[raw=true,float=true,pos=!hb]{Vessel particulars}{tab:vessel-particulars}{}
\small
\setlength{\tabcolsep}{3pt}
\begin{tabular}{@{}p{2.5cm}p{4.2cm}p{0.7cm}@{}}
\hline
\textbf{Parameter} & \textbf{Value} & \textbf{Unit} \\
\hline
Vessel type & Panamax Bulk Carrier & -- \\
Deadweight (DWT) & 75,200 & t \\
Length overall (LOA) & 225.00 & m \\
Length between perpendiculars (LBP) & 217.00 & m \\
Rule length & 213.885 & m \\
Breadth (moulded) & 32.26 & m \\
Depth (moulded) & 19.60 & m \\
Design draught & 12.20 & m \\
Scantling draught & 14.20 & m \\
Block coefficient ($C_b$) & 0.879 & -- \\
Propulsion system & Single low-speed diesel engine, fixed-pitch propeller & -- \\
Specified Maximum Continuous Rating (SMCR) & 8,833 & kW \\
SMCR speed & 105 & rpm \\
Continuous Service Output (CSO) & 7,950 & kW \\
CSO speed & 101.4 & rpm \\
Service speed & 14.5 & kn \\
Trial speed & 15.8 & kn \\
\hline
\end{tabular}
\end{ptable}

The telemetry system integrates measurements from multiple onboard sensors, automation systems, navigation equipment, and external weather data providers. Operational parameters include main engine rotational speed, turbocharger rotational speed, scavenging air pressure, engine load, fuel oil mass flow, and propeller shaft rotational speed, revolutions, torque, power, mechanical energy, and thrust. Additional navigation-related parameters include vessel speed over ground, speed through water, heading, rate of turn, water depth, and fore, aft, and midship drafts. Environmental observations comprise relative and true wind speed and direction, significant wave height, swell height, wave direction and period, sea water temperature, and wind and sea current velocity components. Data acquisition is performed through dedicated shipboard instrumentation, including pressure transducers, revolutions-per-minute (RPM) transducers, torque meters, RPM indicators, the Global Positioning System (GPS), speed log, gyrocompass, echo sounder, the Control Alarm Monitoring System, anemometers, and external weather data providers.
Certain operational and environmental parameters are intentionally represented more than once because they originate from independent onboard instrumentation or external data providers rather than constituting duplicate observations. Specifically, Main Engine Turbocharger Rotational Speed is recorded by the RPM Transducer and the RPM Indicator, respectively. The environmental variables Significant Wave Height, Eastward Sea Water Velocity, and Northward Sea Water Velocity are obtained from two independent weather data providers (Provider MB and Provider S). Retaining these parallel data streams enables cross-validation between independent acquisition systems, supports sensor consistency checks, facilitates anomaly detection and quality assurance during data preprocessing, and allows the selection of the most appropriate signal for subsequent analyses. Consequently, all telemetry streams were preserved in the original dataset to maintain data integrity, maximise transparency, and provide flexibility for subsequent statistical analyses, predictive modelling, and machine learning applications. This approach is consistent with contemporary maritime digitalisation and Internet of Things (IoT) architectures, where redundant sensing and multiple weather data sources are deliberately retained to enhance the reliability, robustness, and resilience of high-frequency operational datasets.

Overall, the dataset comprises 37 high-frequency telemetry variables representing machinery performance, propulsion characteristics, navigation, vessel loading condition, and environmental conditions. The diversity of the recorded variables enables the simultaneous investigation of technical, operational, and environmental factors affecting vessel performance. A complete description of all recorded variables, their corresponding measurement units, and acquisition sources is provided in Table~\ref{tab:data-description}.

\makeatletter
\let\prism@original@dbflt\@dbflt
\def\@dbflt#1{\@ifnextchar[{\@xdblfloat{#1}}{\edef\@tempa{\noexpand\@xdblfloat{#1}[\csname fps@#1\endcsname]}\@tempa}}
\makeatother
\begin{ptable}[raw=true,span=double,pos=!b]{High-frequency telemetry variables included in the dataset.}{tab:data-description}{}
\centering

    \vspace{2mm}

    \footnotesize
    \setlength{\tabcolsep}{2pt}
    \renewcommand{\arraystretch}{0.9}

    \begin{tabularx}{\textwidth}{
        @{}
        p{2.3cm}
        >{\raggedright\arraybackslash}X
        >{\raggedright\arraybackslash}X
        p{1.5cm}
        p{3cm}
        @{}
    }
    \hline
    \textbf{Category} & \textbf{Variable} & \textbf{Description} &
    \textbf{Unit} & \textbf{Source} \\
    \hline

    \multirow{6}{*}{Main Engine}
    & Main Engine Rotational Speed & Main engine crankshaft rotational speed & rpm & Control Alarm Monitoring System \\
    & Main Engine Fuel Load & Main engine load & \% & Fuel Load Transducer \\
    & Main Engine Scavenging Air Pressure & Main engine scavenging air pressure & bar & Pressure Transducer \\
    & Main Engine Turbocharger Rotational Speed & Main engine turbocharger rotational speed & rpm & RPM Transducer \\
    & Main Engine Turbocharger Rotational Speed & Main engine turbocharger rotational speed & rpm & RPM Indicator \\
    & Main Engine Fuel Oil Inlet Mass Flow & Main engine fuel oil inlet mass flow & kg h$^{-1}$ & Torquemeter \\
    \hline

    \multirow{6}{*}{Propulsion System}
    & Propeller Shaft Rotational Speed & Propeller shaft rotational speed & rpm & Torquemeter \\
    & Propeller Shaft Revolutions & Propeller shaft revolutions & rev & Torquemeter \\
    & Propeller Shaft Torque & Propeller shaft torque & N$\cdot$m & Torquemeter \\
    & Propeller Shaft Mechanical Power & Propeller shaft mechanical power & kW & Torquemeter \\
    & Propeller Shaft Mechanical Energy & Propeller shaft mechanical energy & kWh & Torquemeter \\
    & Propeller Shaft Thrust Force & Propeller shaft thrust force & kN & Torquemeter \\
    \hline

    \multirow{5}{*}{Navigation}
    & Speed Over Ground (SOG) & Vessel speed over ground & knots & GPS Receiver \\
    & Speed Through Water (STW) & Vessel speed through water & knots & Speed Log \\
    & True Heading & Vessel true heading & degrees & Gyrocompass \\
    & Rate of Turn & Vessel heading turn rate & deg min$^{-1}$ & Gyrocompass \\
    & Water Depth & Under-keel water depth & m & Echo Sounder \\
    \hline

    \multirow{4}{*}{Loading Condition}
    & Fore Draft & Forward draft & m & Control Alarm Monitoring System \\
    & Aft Draft & Aft draft & m & Control Alarm Monitoring System \\
    & Midship Draft (Port) & Midship draft (port side) & m & Control Alarm Monitoring System \\
    & Midship Draft (Starboard) & Midship draft (starboard side) & m & Control Alarm Monitoring System \\
    \hline

    \multirow{6}{*}{Wind}
    & Relative Wind Speed & Relative wind speed & knots & Anemometer \\
    & Relative Wind Angle & Relative wind direction & degrees & Anemometer \\
    & True Wind Speed & True wind speed & knots & Weather Provider (MB) \\
    & True Wind Angle & True wind direction & degrees & Weather Provider (MB) \\
    & Eastward Wind Velocity & Eastward wind velocity & m s$^{-1}$ & Weather Provider (S) \\
    & Northward Wind Velocity & Northward wind velocity & m s$^{-1}$ & Weather Provider (S) \\
    \hline

    \multirow{10}{*}{Waves \& Sea}
    & Significant Wave Height & Significant wave height & m & Weather Provider (MB) \\
    & Significant Wave Height & Significant wave height & m & Weather Provider (S) \\
    & Swell Significant Height & Significant swell height & m & Weather Provider (MB) \\
    & Wave Direction & True wave direction & degrees & Weather Provider (S) \\
    & Wave Period & Wave period & s & Weather Provider (S) \\
    & Sea Water Temperature & Sea surface temperature & \textdegree C & Weather Provider (S) \\
    & Eastward Sea Water Velocity & Eastward sea current velocity & m s$^{-1}$ & Weather Provider (MB) \\
    & Northward Sea Water Velocity & Northward sea current velocity & m s$^{-1}$ & Weather Provider (MB) \\
    & Eastward Sea Water Velocity & Eastward sea current velocity & m s$^{-1}$ & Weather Provider (S) \\
    & Northward Sea Water Velocity & Northward sea current velocity & m s$^{-1}$ & Weather Provider (S) \\
    \hline

    \end{tabularx}
\end{ptable}
\makeatletter
\let\@dbflt\prism@original@dbflt
\makeatother

\section{Experimental Environment}
\label{sec:experimental_environment}
{This section describes the software and computational environments used to implement and evaluate the proposed classical and QML models. Although the methodology presented in Section~\ref{sec:methodology} is independent of any specific software platform or hardware architecture, the experiments reported in this work were conducted using dedicated classical and quantum computing environments. The classical environment is described in Section~\ref{subsec:classical_environment}, while the quantum software and infrastructure are detailed in Section~\ref{subsec:quantum_environment}.}

\subsection{Classical Environment}
\label{subsec:classical_environment}
{The classical machine learning models considered in this study were implemented in Python using the \texttt{scikit-learn} framework. All experiments were conducted using the same datasets, preprocessing procedures, training-validation splits, and performance evaluation metrics employed for the quantum models in order to ensure fair and consistent comparisons.}

{Classical SVR models were configured according to the hyperparameter optimisation strategy described in Section~\ref{subsec:hyperparameter-tuning}. Additional data processing and numerical computations were performed using the NumPy and Pandas libraries. This environment served as the baseline for assessing the predictive performance of the proposed QSVR configurations.}

\subsection{Quantum Environment}
\label{subsec:quantum_environment}

{This subsection describes the software and computational infrastructure used to implement and evaluate the proposed QSVR models. The quantum software framework is introduced in Section~\ref{subsubsec:quantum-framework}, the custom QSVR library is presented in Section~\ref{subsubsec:quantum-library}, and the emulator and hardware configurations are described in Sections~\ref{subsubsec:quantum-emulator} and~\ref{subsubsec:hardware-infrastructure}, respectively.}

\subsubsection{Quantum Software Framework}
\label{subsubsec:quantum-framework}

{The QML models developed in this work were implemented using the Qaptiva software ecosystem \citep{qaptiva800}. This environment supports quantum circuit construction, state manipulation, and hybrid quantum-classical workflows through a unified programming interface. Its abstraction of the execution backend enables the same algorithmic implementation to be used across different execution environments.}

{Within this framework, Qaptiva was used to implement the quantum feature maps considered in this study, generate quantum states, construct quantum kernels, and support the execution of QSVR experiments.}

\subsubsection{Custom QSVR Library}
\label{subsubsec:quantum-library}
{As Qaptiva does not provide a native implementation of QSVR, a dedicated software library was developed to support the experimentation performed in this work.}

{The library integrates the quantum feature maps analysed in this study, described in Sections~\ref{subsubsec:angle-encoding} and~\ref{subsubsec:zzfeaturemap}, and provides the functionality required for quantum state generation, fidelity computation, kernel matrix construction, and integration with classical SVR solvers. Additionally, utilities for experiment management, automated kernel generation, and large-scale experimental execution were incorporated into the framework. Quantum kernel entries are computed exactly via statevector emulation rather than shot-based sampling, eliminating measurement noise at the cost of classical simulability. To reduce computational overhead, statevectors are cached and the full kernel matrix is constructed via a single matrix multiplication over the cached statevectors, avoiding redundant statevector-emulator executions for repeated data points.}

{To ensure reproducibility and facilitate systematic benchmarking, all experiments are defined through JavaScript Object Notation (JSON) configuration files. These files specify the selected feature map, circuit repetitions, encoding parameters, QSVR hyperparameters, and additional execution settings. This modular architecture simplifies experiment management and enables the straightforward incorporation of new feature maps and kernel generation techniques in future developments.}

{Once generated, the quantum kernel matrix is supplied to a classical SVR optimisation procedure, ensuring that the differences observed between classical and quantum configurations arise exclusively from the kernel construction process.}

\subsubsection{Quantum Emulator Configuration}
\label{subsubsec:quantum-emulator}

{Quantum kernel evaluation was performed using the LinalgQPU backend available within the Qaptiva platform \citep{qaptiva_linalg}. This backend relies on a full statevector representation, enabling exact emulation of quantum circuit evolution and direct access to the amplitudes of the generated quantum states.}

{Unlike real quantum hardware or shot-based emulators, statevector emulation is not affected by measurement noise, finite sampling effects, or hardware-induced errors. Consequently, the fidelity values used to compute the entries of the quantum kernel correspond to exact overlaps between quantum states.}

{The adoption of exact statevector emulation allows the experimental analysis to isolate and evaluate exclusively the influence of the selected feature maps and kernel constructions. Therefore, any differences observed among the evaluated QSVR configurations originate from the characteristics of the encoding schemes rather than from hardware noise or stochastic sampling uncertainty.}

{All quantum experiments reported in this study were conducted using this exact emulation strategy.}

\subsubsection{Hardware Infrastructure}
\label{subsubsec:hardware-infrastructure}

{Classical and quantum experiments were executed on the Qaptiva 802 platform, configured with a 1.92~TB E1S storage device, 2~TB of random access memory (RAM), and two 28-core processors operating at 2~GHz \citep{qaptiva800}. The infrastructure does not include graphics processing units (GPUs); therefore, all quantum emulations were performed exclusively on its central processing units (CPUs). All development and experiments were carried out using Python version 3.12.9.}

{It should be noted that statevector-based quantum emulation requires substantially larger memory resources than conventional machine learning workloads, since the dimensionality of the quantum state space increases exponentially with the number of qubits. Consequently, memory availability represents a critical factor for the execution of large-scale quantum kernel simulations. The detailed description of both execution environments contributes to the reproducibility of the reported results and facilitates future comparisons with related studies.}

\section{Results}
\label{sec:results}

After exploratory data analysis of the raw data and before implementing the classical and quantum SVR models, data preprocessing was conducted. The first step was steady-state identification, which aimed to discard instances that did not correspond to operational steady states. Following a heuristic evaluation, the window size was set to 27 instances and the minimum steady-state length to 270 instances. Shorter steady-state runs were removed to avoid including spurious segments arising near transients. After applying these steps, 14,049 instances were retained for the subsequent analysis.

The configuration names used throughout this section correspond to the rows in Table~\ref{tab:quantum-configurations}: AngleEncoding\_1 denotes Angle Encoding with a linear entanglement topology, and AngleEncoding\_2 denotes the full topology. ZZFeatureMap\_1 and ZZFeatureMap\_2 denote the configurations with $L=1$ and $L=2$ repetitions, respectively.

The results are organised into three experimental stages:
\begin{enumerate}
    \item \textit{Exploratory analysis with three input features}: examines how dataset size and missingness affect imputation performance and overfitting, rather than seeking a definitive best model. Each model was evaluated across all combinations of dataset size and missingness. Dataset size ranged from 10\% to 100\% of the full dataset in 10-percentage-point increments, while the missing-target fraction ranged from 10\% to 90\%, also in 10-percentage-point increments. Each configuration was repeated using 10 independently generated random missingness splits, and the reported metrics were averaged across these repetitions.
    \item \textit{Scalability analysis with increasing feature dimensionality}: uses 100\% of the dataset with 90\% missing target values to investigate how increasing the number of features and qubits affects predictive performance and computational cost under the highest missingness rate considered.
    \item \textit{Comparison under the selected representative configuration}: combines the findings from the first two stages in a comparison using 14 features, 60\% of the dataset, and the same 90\% missingness rate. This resource-conscious compromise reflects the limited and model-dependent improvements from adding features in the full-dataset scalability analysis and the sufficiently low overfitting observed with a reduced dataset in the exploratory analysis.
\end{enumerate}

\subsection{Exploratory analysis with three input features}
\label{subsec:results-exploratory}

Rather than selecting these variables solely on the basis of their statistical association with fuel flow, the selection was guided by the objective of representing three principal physical drivers of fuel consumption: propulsion loading, hull/hydrostatic condition, and environmental resistance. One representative variable was selected for each category: main engine RPM, mean draft, and relative wind speed, respectively. This selection was also consistent with other studies in fuel consumption modelling \citep{GKEREKOS2019106282}. Each quantum model therefore used three qubits under the adopted feature-to-qubit encoding. The purpose of this stage was to characterise sensitivity to data availability and identify a suitable scenario for further analysis, not to optimise the feature set or establish a final model ranking.

The impact of dataset size is also assessed. To do so, the dataset is randomly partitioned into subsets of varying size, ranging from 10\% to 100\% of the total dataset in increments of 10\%. For each partition, missing values are injected completely at random. Specifically, nine MCAR scenarios are considered, with missing values introduced at rates ranging from 10\% to 90\% in increments of 10\%. The instances with observed target values in each subset are used for training, while the artificially masked target values are used to evaluate imputation performance. Table~\ref{tab:train-test} reports the resulting training and test set sizes for each missing-data percentage using all 14,049 preprocessed instances. Finally, the data were prepared for cross-validation during hyperparameter optimisation, and normalisation was applied.

\begin{ptable}[raw=true,float=true,pos=htbp]{Training and test set sizes for all 14,049 preprocessed instances at each MCAR missing-target percentage.}{tab:train-test}{}
\renewcommand{\arraystretch}{1.15}
\begin{tabular}{lrr}
\toprule
\textbf{Missing values} & \textbf{Train size} & \textbf{Test size} \\
\midrule
10\% & 12,644 & 1,405 \\
20\% & 11,239 & 2,810 \\
30\% & 9,834  & 4,215 \\
40\% & 8,429  & 5,620 \\
50\% & 7,025  & 7,024 \\
60\% & 5,620  & 8,429 \\
70\% & 4,215  & 9,834 \\
80\% & 2,810  & 11,239 \\
90\% & 1,405  & 12,644 \\
\bottomrule
\end{tabular}
\end{ptable}

Table~\ref{tab:overall_summary} summarises the aggregated performance of the four classical kernels (RBF, polynomial, linear, and sigmoid) and the four quantum configurations (ZZFeatureMap\_1, ZZFeatureMap\_2, AngleEncoding\_1, and AngleEncoding\_2) across all dataset sizes and missing-data fractions. The evaluation metrics are defined in Section~\ref{subsec:evaluation}. Quantum execution times correspond to statevector-emulation times and are provided only as implementation-dependent reference values.

Overall, all methods achieved strong predictive performance ($R^2 > 0.91$), indicating that both classical and quantum approaches are capable of accurately modelling fuel-consumption rates. ZZFeatureMap\_1 delivered the best aggregate results, achieving the highest mean test $R^2$ and the lowest imputation errors. However, its advantage over the classical RBF kernel was small and generally within the observed experimental variability. A second performance group, comprising ZZFeatureMap\_2, the polynomial kernel, AngleEncoding\_1, and AngleEncoding\_2, exhibited very similar imputation performance. In contrast, the linear and sigmoid kernels consistently produced lower $R^2$ values and higher imputation errors, suggesting a more limited ability to capture the nonlinear relationships present in the data.

Regarding generalisation, all methods showed low relative overfitting, with classical kernels exhibiting slightly lower values on average. Computational costs varied with the model and data-availability scenario. The mean-based timing summary in Table~\ref{tab:overall_summary} is sensitive to isolated runtime peaks attributed to transient computational-infrastructure overhead. Figure~\ref{fig:comparison-runtime} therefore reports median runtimes to characterise typical execution costs more robustly. Overall, the results indicate that quantum feature maps can achieve imputation performance comparable to that of the best-performing classical kernels for the fuel-consumption estimation problem considered.

\begin{ptable}[raw=true,float=true,pos=htbp]{Exploratory performance with three input features (three qubits for quantum models) over all dataset sizes, missing fractions and iterations; cells are mean $\pm$ standard deviation. Metrics are defined in Section~\ref{subsec:evaluation}; the best value in each column is in bold; quantum times are statevector-emulation runtimes, not hardware ones.}{tab:overall_summary}{}
\begin{adjustbox}{max width=\textwidth}
\begin{tabular}{llllll}
\toprule
Method & Mean MAE (kg h$^{-1}$) & Mean RMSE (kg h$^{-1}$) & Mean $R^2$ & Mean Rel.\ overfit & Mean runtime (s) \\
\midrule
rbf & 33.65 $\pm$ 1.76 & 49.44 $\pm$ 2.24 & 0.944 $\pm$ 0.005 & 0.003 $\pm$ 0.007 & 15.0 $\pm$ 18.2 \\
poly & 35.73 $\pm$ 1.27 & 51.25 $\pm$ 1.65 & 0.940 $\pm$ 0.004 & 0.002 $\pm$ 0.006 & 198.5 $\pm$ 244.2 \\
linear & 41.08 $\pm$ 0.98 & 59.45 $\pm$ 1.64 & 0.919 $\pm$ 0.004 & \bestvalue{0.000 $\pm$ 0.008} & \bestvalue{3.4 $\pm$ 4.1} \\
sigmoid & 41.14 $\pm$ 1.02 & 59.58 $\pm$ 1.64 & 0.918 $\pm$ 0.004 & \bestvalue{0.000 $\pm$ 0.008} & 21.9 $\pm$ 27.7 \\
ZZFeatureMap\_1 & \bestvalue{32.59 $\pm$ 1.63} & \bestvalue{48.60 $\pm$ 2.85} & \bestvalue{0.945 $\pm$ 0.007} & 0.006 $\pm$ 0.009 & 25.6 $\pm$ 15.7 \\
ZZFeatureMap\_2 & 33.76 $\pm$ 3.10 & 50.43 $\pm$ 4.93 & 0.941 $\pm$ 0.014 & 0.008 $\pm$ 0.015 & 29.2 $\pm$ 18.1 \\
AngleEncoding\_1 & 34.51 $\pm$ 1.13 & 51.47 $\pm$ 1.66 & 0.939 $\pm$ 0.004 & 0.002 $\pm$ 0.006 & 27.5 $\pm$ 42.2 \\
AngleEncoding\_2 & 35.05 $\pm$ 1.18 & 51.77 $\pm$ 1.72 & 0.938 $\pm$ 0.004 & 0.002 $\pm$ 0.006 & 25.6 $\pm$ 33.8 \\
\bottomrule
\end{tabular}
\end{adjustbox}
\end{ptable}

This section retains the RMSE plot as the main summary of predictive error, alongside the imputed-versus-real scatter plots, the relative-overfitting plot, and the runtime plot. Because MAE, RMSE, and $R^2$ show the same overall performance pattern, RMSE is retained as the representative error measure, since it reports error in the target's physical units (kg~h$^{-1}$) and gives greater weight to larger deviations. The MAE and $R^2$ plots and their discussion are provided in Appendix~\ref{app:exploratory-additional}. The same reason applies to relative  and absolute overfitting, keeping the relative one in this case. The absolute generalisation-gap plot and its discussion are also included in Appendix~\ref{app:exploratory-additional}.

From an application-oriented perspective, this analysis is particularly relevant for fuel-consumption modelling, where both the amount of available operational data and the proportion of missing observations may vary substantially across vessels, voyages, and onboard monitoring systems. Consequently, evaluating model behaviour under different levels of data availability is essential for assessing the practical applicability and robustness of the proposed approaches in real-world maritime environments.

Figure~\ref{fig:comparison-rmse} shows that ZZFeatureMap\_1 and RBF consistently achieve the lowest imputation errors across most scenarios. ZZFeatureMap\_2, the polynomial kernel, AngleEncoding\_1, and AngleEncoding\_2 exhibit comparable RMSE values, whereas the linear and sigmoid kernels form the highest-error group. Error levels increase moderately with the fraction of missing data, this gap is more relevant for the smallest dataset partitions. For dataset sizes above 40--50\%, most methods exhibit stable RMSE values and reduced variability, while ZZFeatureMap\_2 shows the strongest degradation and widest uncertainty bands under data-scarce conditions.

\begin{pfig}[float=false,span=double]{Mean RMSE on the imputed test set across dataset sizes and missing-data percentages.}{fig:comparison-rmse}
\centering
\includegraphics[width=.9\textwidth,height=.74\textheight,keepaspectratio]{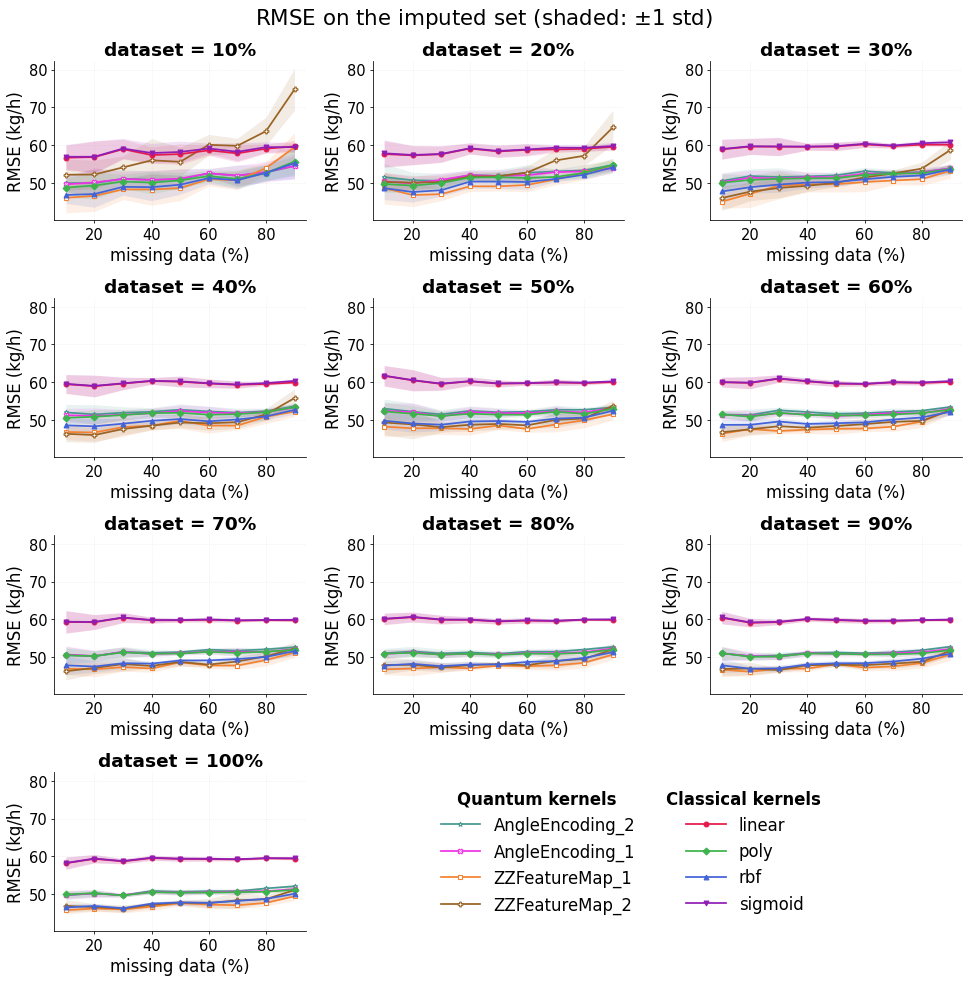}
\end{pfig}

Figures~\ref{fig:imputation-scatter-easy}, \ref{fig:imputation-scatter-intermediate}, and \ref{fig:imputation-scatter-extreme} illustrate the relationship between observed and imputed fuel-flow values under favourable, intermediate, and highly constrained data-availability conditions, respectively. Across all scenarios, both classical and quantum approaches preserve a clear positive correlation between observed and imputed values, indicating that the underlying structure of the target variable is consistently captured.

\begin{pfig}[float=false,span=double]{Imputed versus real main-engine fuel flow for all classical SVR and QSVR configurations using 100\% of the dataset and 10\% missing data. The dashed diagonal represents perfect agreement.}{fig:imputation-scatter-easy}
\centering
\includegraphics[width=0.9\textwidth,height=\textheight,keepaspectratio]{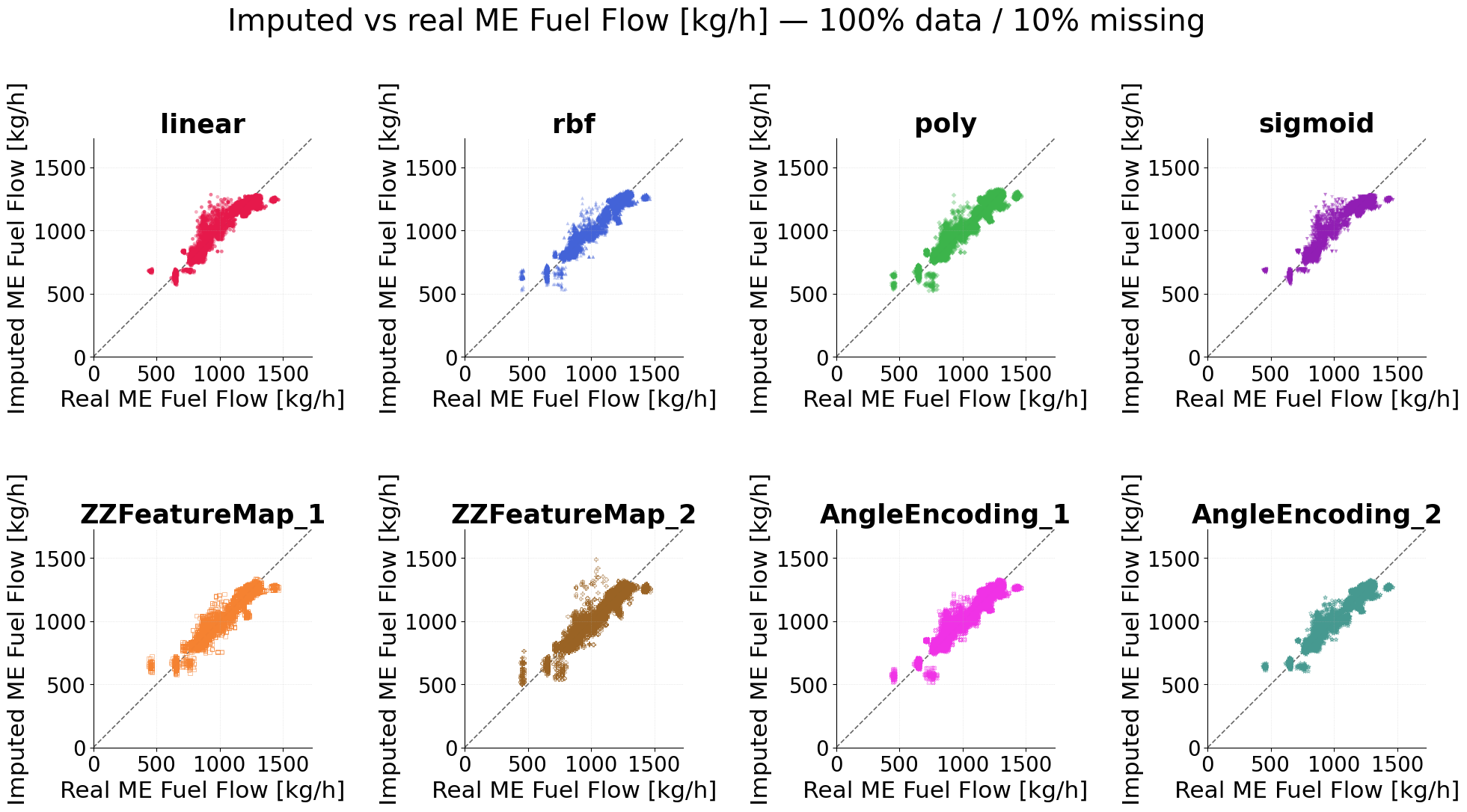}
\end{pfig}

\begin{pfig}[float=false,span=double]{Imputed versus real main-engine fuel flow for all classical SVR and QSVR configurations using 50\% of the dataset and 50\% missing data. The dashed diagonal represents perfect agreement.}{fig:imputation-scatter-intermediate}
\centering
\includegraphics[width=.9\textwidth,height=.74\textheight,keepaspectratio]{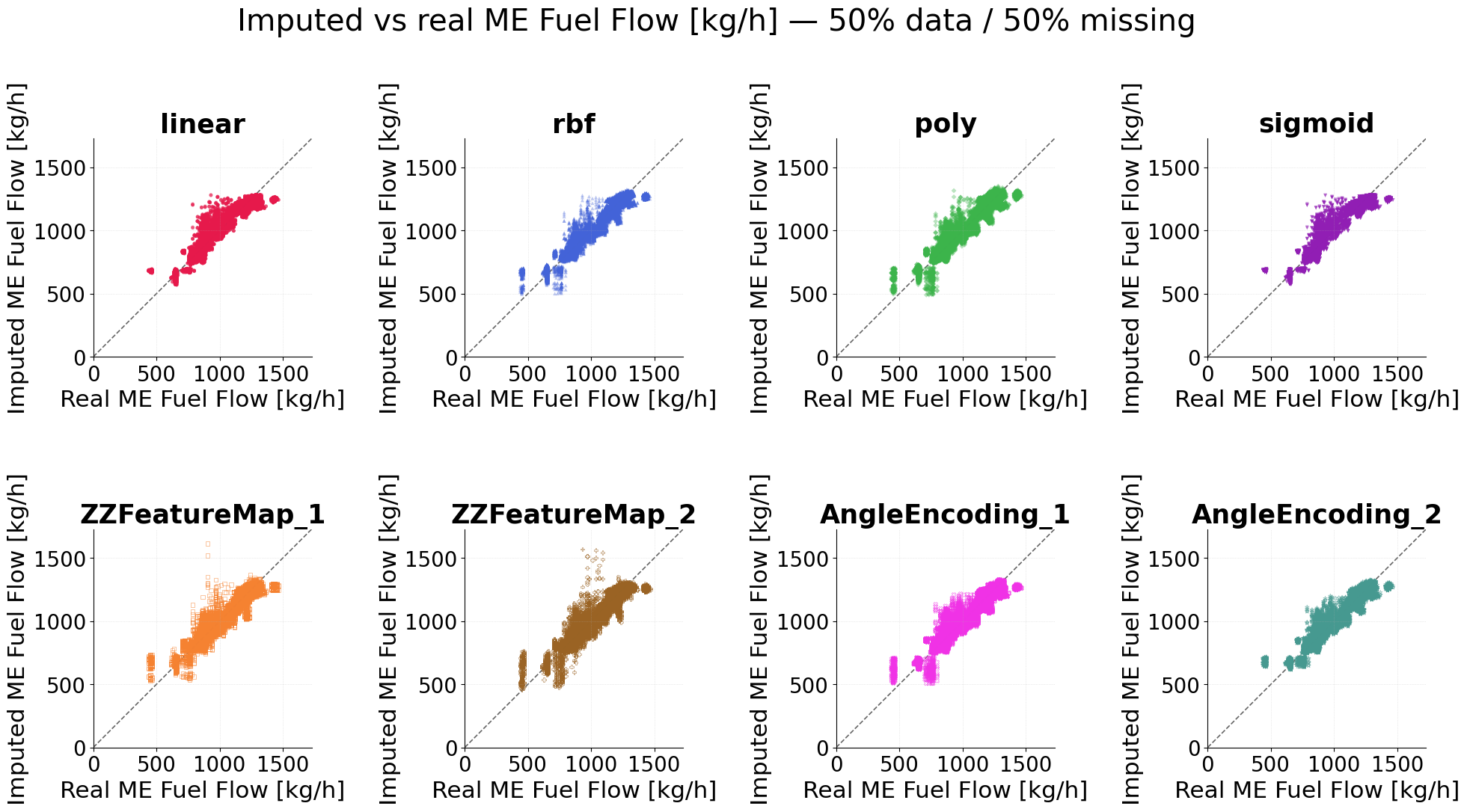}
\end{pfig}

\begin{pfig}[float=false,span=double]{Imputed versus real main-engine fuel flow for all classical SVR and QSVR configurations using 10\% of the dataset and 90\% missing data. The dashed diagonal represents perfect agreement.}{fig:imputation-scatter-extreme}
\centering
\includegraphics[width=.9\textwidth,height=.74\textheight,keepaspectratio]{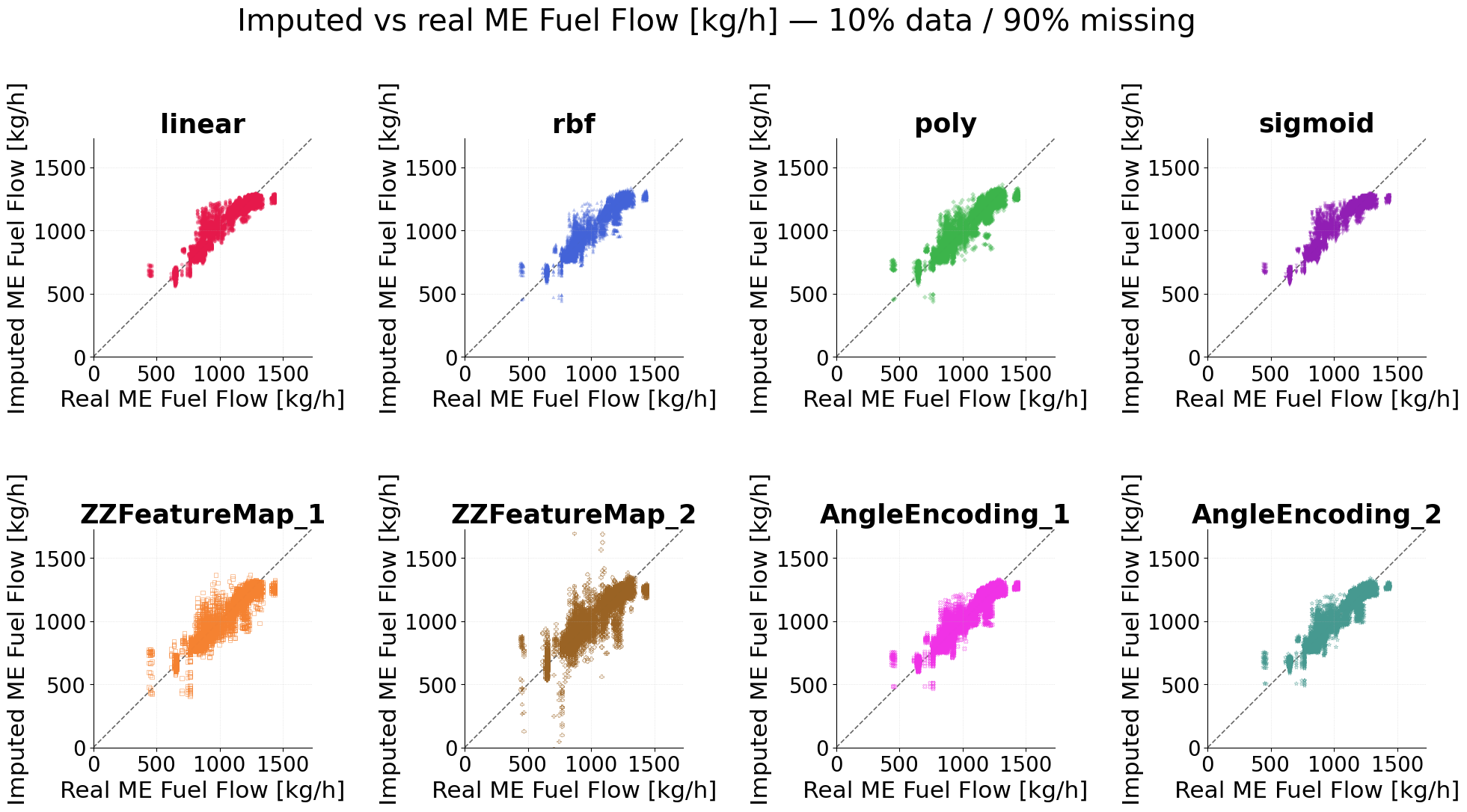}
\end{pfig}

Under favourable conditions (Figure~\ref{fig:imputation-scatter-easy}), most methods produce predictions closely aligned with the identity line, with the RBF kernel, ZZFeatureMap\_1, the polynomial kernel, AngleEncoding\_1, and AngleEncoding\_2 exhibiting the most compact point clouds. As data availability decreases (Figure~\ref{fig:imputation-scatter-intermediate}), a moderate increase in dispersion becomes apparent, although the overall predictive behaviour remains largely unchanged. In the most challenging scenario (Figure~\ref{fig:imputation-scatter-extreme}), the spread around the identity line increases further, particularly for ZZFeatureMap\_2, which exhibits the largest deviations and variability.

A common pattern across all scenarios is the compression of the prediction range, leading to slight overestimation of low fuel-flow values and underestimation of high values. This regression-to-the-mean effect becomes more pronounced as dataset size decreases and the proportion of missing data increases. Nevertheless, even under severe data scarcity, the strongest-performing methods, notably RBF, ZZFeatureMap\_1, AngleEncoding\_1, and AngleEncoding\_2, maintain a relatively tight alignment with the diagonal, demonstrating robust predictive capability. These visual observations are consistent with the quantitative results reported in Table~\ref{tab:overall_summary} and Figure~\ref{fig:comparison-rmse}.

Figure~\ref{fig:comparison-overfitting} indicates that relative overfitting is generally negligible across all methods, particularly for dataset sizes above 30--40\%, where values remain close to zero regardless of the missing-data fraction. The main exception is ZZFeatureMap\_2, which exhibits increased sensitivity under data-scarce conditions, especially at high missing-data levels. ZZFeatureMap\_1 shows a similar but less pronounced behaviour. Small negative values are attributable to statistical variability between random repetitions rather than systematic test-set superiority.

\begin{pfig}[float=false,span=double]{Relative overfitting across dataset sizes and missing-data percentages.}{fig:comparison-overfitting}
\centering
\includegraphics[width=0.9\textwidth,height=\textheight,keepaspectratio]{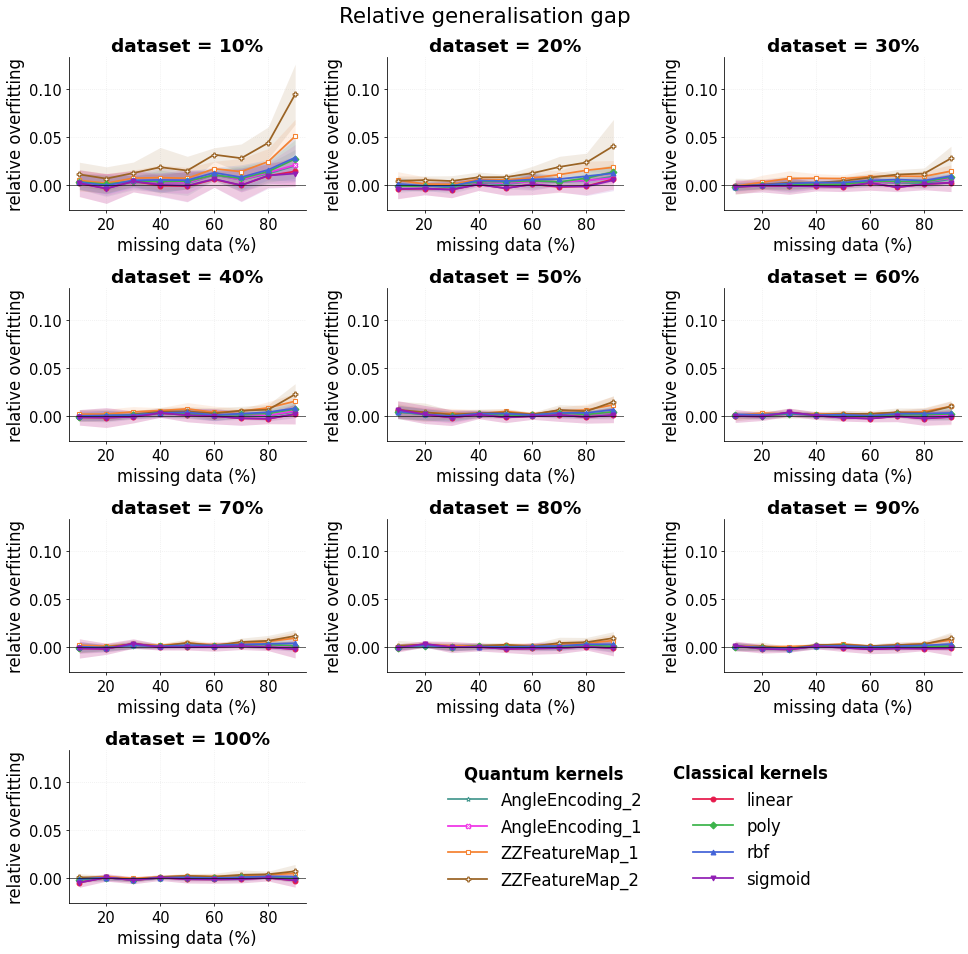}
\end{pfig}

Figure~\ref{fig:comparison-runtime} reports median runtimes over repeated iterations rather than means. This choice reduces the influence of peaks observed in individual runs, which were attributed to transient overhead in the computational infrastructure at the time of execution. Median runtime generally increases with dataset size and decreases as the fraction of missing data grows, since fewer observed samples remain for training. The linear kernel is consistently the fastest method, while the polynomial kernel usually incurs the highest cost, particularly for large datasets with low missing-data percentages. RBF and sigmoid show a stronger decrease in runtime with increasing missingness than the quantum configurations, whose median runtimes are comparatively stable within each dataset size.

\begin{pfig}[float=false,span=double]{Median runtime over repeated iterations by dataset size and missing-data percentage. The median reduces the influence of isolated timing peaks attributed to transient computational-infrastructure overhead. The vertical axis uses a logarithmic scale. Quantum times correspond to statevector emulation.}{fig:comparison-runtime}
\centering
\includegraphics[width=0.9\textwidth,height=\textheight,keepaspectratio]{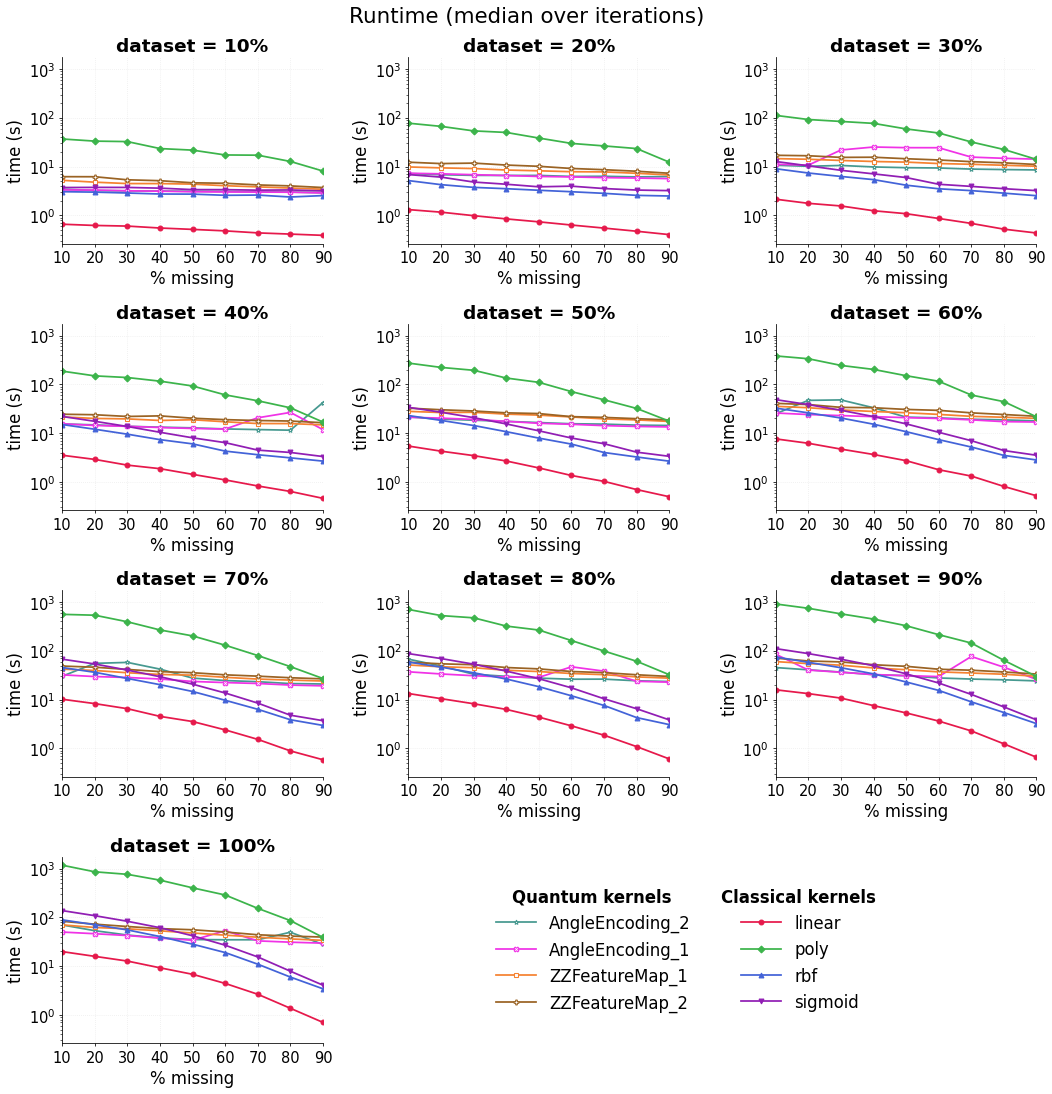}
\end{pfig}

Taken together, the exploratory results support the use of a reduced dataset for the final comparison. At 60\% of the available instances, predictive performance remains broadly comparable to that obtained with larger subsets, while overfitting is sufficiently low for that comparison. Before adopting this reduced-data configuration, however, the scalability analysis performed in the next section uses the full dataset with 90\% missing target values. This is the highest missingness rate investigated and provides a demanding setting for assessing the effect of feature dimensionality while keeping the dataset size fixed. The 60\% subset is reserved for the final stage, rather than used in the scalability tests.

\subsection{Scalability analysis with increasing feature dimensionality}
\label{subsec:results-scalability}

The second stage extends the exploratory analysis by varying the number of input features from 1 to 18, using 100\% of the preprocessed dataset and masking 90\% of the target values under MCAR.

The predictor variables used for fuel-flow imputation are ordered according to their expected relevance to main-engine fuel consumption, considering both the physical relationships governing marine propulsion and the incremental operational information represented by each parameter. Priority was given to variables directly characterising engine operating state and propulsion demand, including engine load, shaft power, engine rotational speed, scavenging air pressure, shaft torque, and turbocharger operating conditions. These parameters are closely associated with the power developed by the main engine and consequently with instantaneous fuel demand. Navigation and loading variables, such as speed through water, speed over ground, and vessel draft, were subsequently considered because they characterise the vessel's operating condition and the hydrodynamic response associated with the required propulsion power. Environmental variables were assigned lower priority because wind, waves, and currents affect fuel consumption primarily indirectly, through their influence on vessel resistance and the propulsion power required to maintain a given operating condition.

The ordering also considered the incremental information provided by each variable. Variables representing closely related or parallel measurements are not necessarily assigned consecutive positions, since their simultaneous inclusion may provide limited additional information despite strong individual relationships with fuel consumption. This consideration is particularly relevant for measurements obtained from independent sensors or weather providers representing the same physical quantity. The resulting hierarchy therefore reflects a combination of physical relevance, expected predictive capability, measurement independence, and incremental information content. For the scalability analysis, predictors were introduced sequentially according to this ordering, allowing model performance to be evaluated as the dimensionality of the input space and corresponding number of qubits increased. Table~\ref{tab:feature_ranking} shows the selected features ranked by their relevance to fuel flow, along with the rationale for each selection.

\begin{ptable}[raw=true,span=double,pos=t]{Ranking of input variables according to their expected relevance to main engine fuel flow.}{tab:feature_ranking}{}
\centering

\small
\setlength{\tabcolsep}{3pt}
\begin{tabularx}{\textwidth}{@{}c >{\raggedright\arraybackslash}p{0.28\textwidth} >{\raggedright\arraybackslash}X@{}}
\hline
\textbf{Rank} & \textbf{Variable} & \textbf{Reason for relevance to fuel flow} \\
\hline
1  & Main Engine Fuel Load (\%) & Most direct representation of engine operating demand and therefore expected fuel requirement \\
2  & Propeller Shaft Mechanical Power (kW) & Direct measure of delivered propulsion power, strongly physically related to fuel consumption \\
3  & Propeller Shaft Torque (N$\cdot$m) & Directly associated with engine/shaft loading and delivered power \\
4  & Main Engine Rotational Speed (rpm) & Fundamental engine operating parameter governing power and fuel demand \\
5  & Main Engine Scavenging Air Pressure (bar) & Strong indicator of engine load and combustion operating condition \\
6  & Main Engine Turbocharger Rotational Speed, RPM Transducer & Closely responds to engine load, exhaust energy, and combustion demand \\
7  & Propeller Shaft Rotational Speed (rpm) & Closely linked to engine operating point and propulsion demand \\
8  & Main Engine Turbocharger Rotational Speed, RPM Indicator & Same physical relevance as the other turbocharger measurement, but lower incremental information because it measures the same quantity independently \\
9  & Speed Through Water & Represents vessel hydrodynamic response to propulsion and is more directly related to propulsion demand than SOG \\
10 & Speed Over Ground & Strong operational indicator, but affected by current and therefore less directly connected to engine demand than STW \\
11 & Propeller Shaft Thrust Force (kN) & Represents propulsion output and resistance demand \\
12 & Aft Draft & Loading condition influences resistance and propulsion requirement \\
13 & Fore Draft & Same general relationship through displacement and vessel resistance \\
14 & Midship Draft (Port) & Indicator of vessel loading/displacement \\
15 & Midship Draft (Starboard) & Similar information to port midship draft, therefore limited additional information \\
16 & Relative Wind Speed & Directly represents wind experienced by the vessel and resulting aerodynamic resistance \\
17 & Relative Wind Angle & Determines the directionality of wind resistance \\
18 & Significant Wave Height, Provider MB & Sea state influences added resistance and propulsion demand \\
\hline
\end{tabularx}
\end{ptable}

All 14,049 instances are included in the partition, but only the 1,405 instances with observed target values are used for training; the remaining 12,644 instances are used to evaluate imputation. In the quantum models, each feature is encoded using one qubit, so increasing feature dimensionality also increases the number of qubits. The objective is to assess whether the predictive gains from richer input representations justify extending the computationally more demanding statevector-based experiments. Figures~\ref{fig:scalability-rmse}--\ref{fig:scalability-max-error} report RMSE, $R^2$, and maximum absolute error across feature dimensionalities for the linear, RBF and sigmoid kernels and the four quantum feature maps (the polynomial kernel is not shown). For readability, the imputed-versus-real scatter plots are limited to the RBF kernel and AngleEncoding\_1, selected as representative classical and quantum models, respectively; for each method, panel A covers one to nine features and panel B covers 10 to 18 features.

Figure~\ref{fig:scalability-rmse} shows that RMSE improves most as the number of features increases from one to around six. Beyond that point, the reduction becomes more gradual and non-monotonic, and the curves do not show a uniform improvement with every additional feature. The linear and sigmoid kernels level off from eight features onwards at clearly higher errors, whereas the lowest RMSE values are obtained between 14 and 17 features. Both ZZFeatureMap configurations deteriorate beyond 14 features, more markedly for L = 2. This behaviour is consistent with the concentration of the off-diagonal quantum kernel values discussed in \ref{subsubsec:quantum-configurations}, which becomes more pronounced as the number of qubits and the circuit expressivity increase \citep{Thanasilp2024}. Confirming this mechanism would, however, require an analysis of the kernel-value distribution.

\begin{pfig}[float=false,span=single]{RMSE versus number of input features for the linear, RBF and sigmoid kernels and the quantum feature maps in the scalability analysis.}{fig:scalability-rmse}
\centering
\includegraphics[width=.75\textwidth,height=.68\textheight,keepaspectratio]{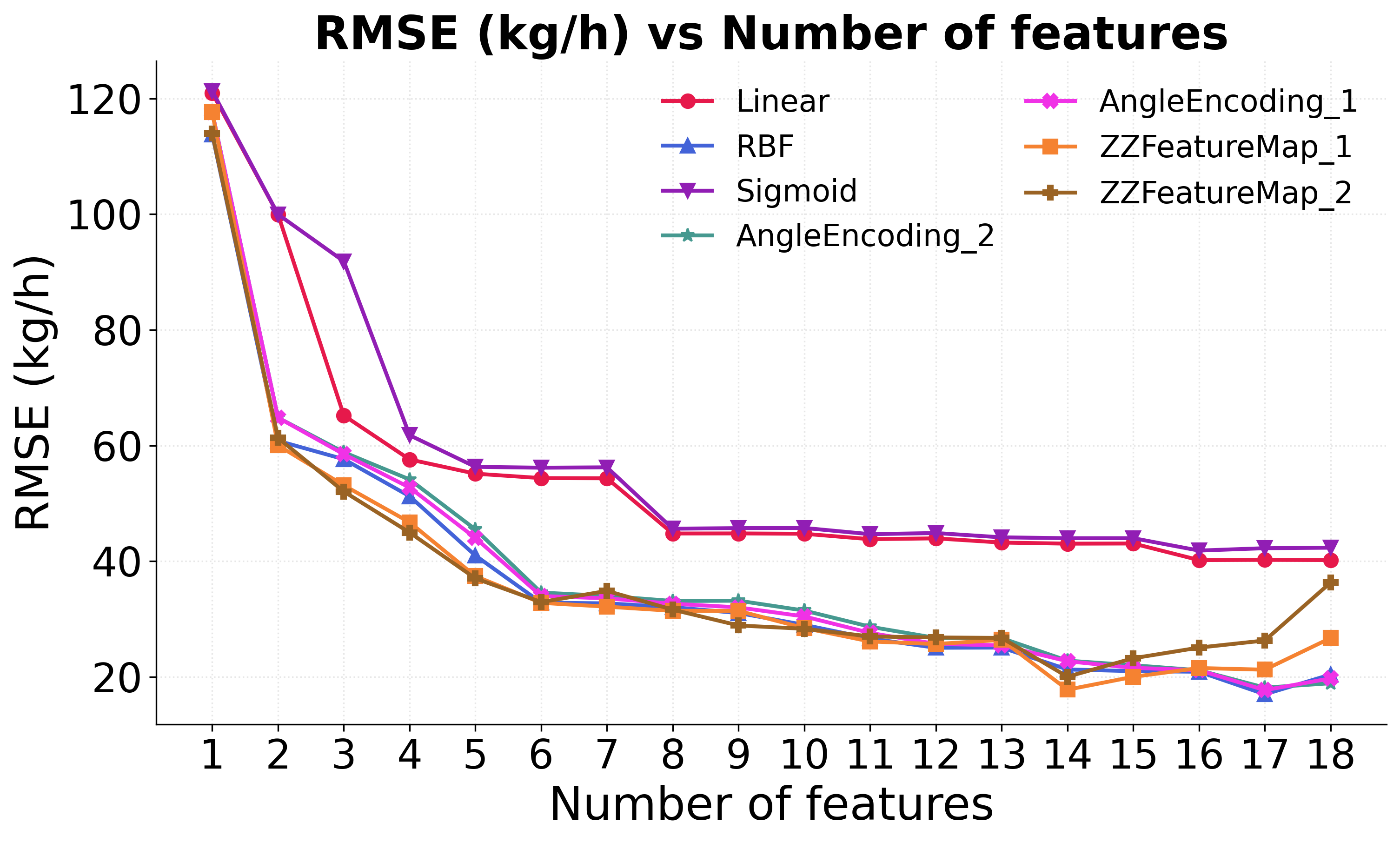}
\end{pfig}

Figure~\ref{fig:scalability-r2} shows that $R^2$ increases steeply up to about six features and only marginally thereafter. With a single feature, $R^2$ lies between 0.66 and 0.70, whereas from eight features onwards all methods reach values of about 0.95 or higher, with some local variation across feature maps.

\begin{pfig}[float=false,span=single]{$R^2$ versus number of input features for the linear, RBF and sigmoid kernels and the quantum feature maps in the scalability analysis.}{fig:scalability-r2}
\centering
\includegraphics[width=.75\textwidth,height=.68\textheight,keepaspectratio]{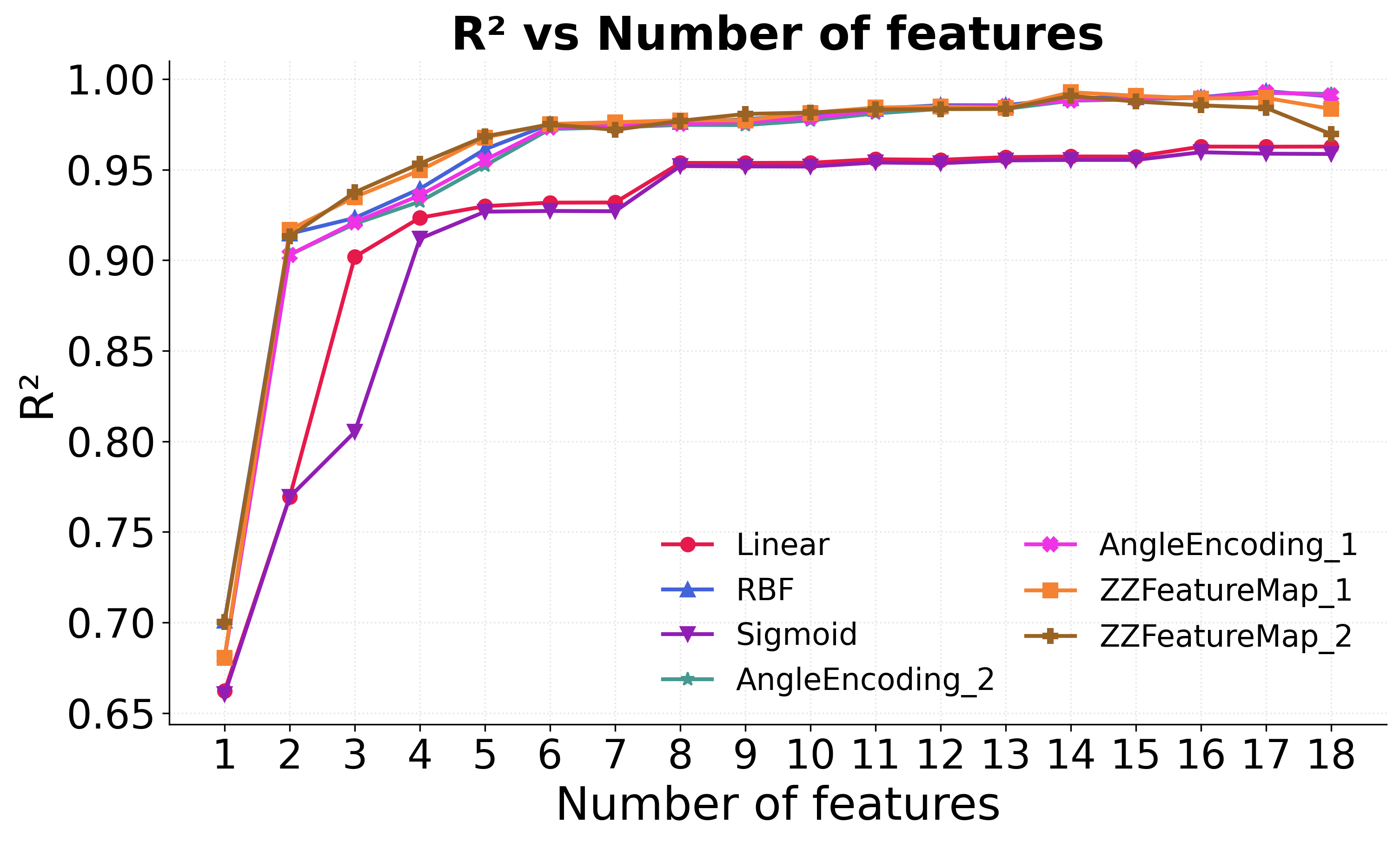}
\end{pfig}

Figure~\ref{fig:scalability-max-error} shows a more irregular pattern, with isolated peaks at several feature dimensionalities. Since maximum absolute error reflects only the single largest prediction deviation, these peaks do not necessarily correspond to changes in typical error captured by RMSE. Taken together, the three metrics support treating 14 features as a practical accuracy--resource compromise for this case study, rather than as a uniquely optimal dimensionality.

\begin{pfig}[float=false,span=single]{Maximum absolute error versus number of input features for the linear, RBF and sigmoid kernels and the quantum feature maps in the scalability analysis.}{fig:scalability-max-error}
\centering
\includegraphics[width=.75\textwidth,height=.68\textheight,keepaspectratio]{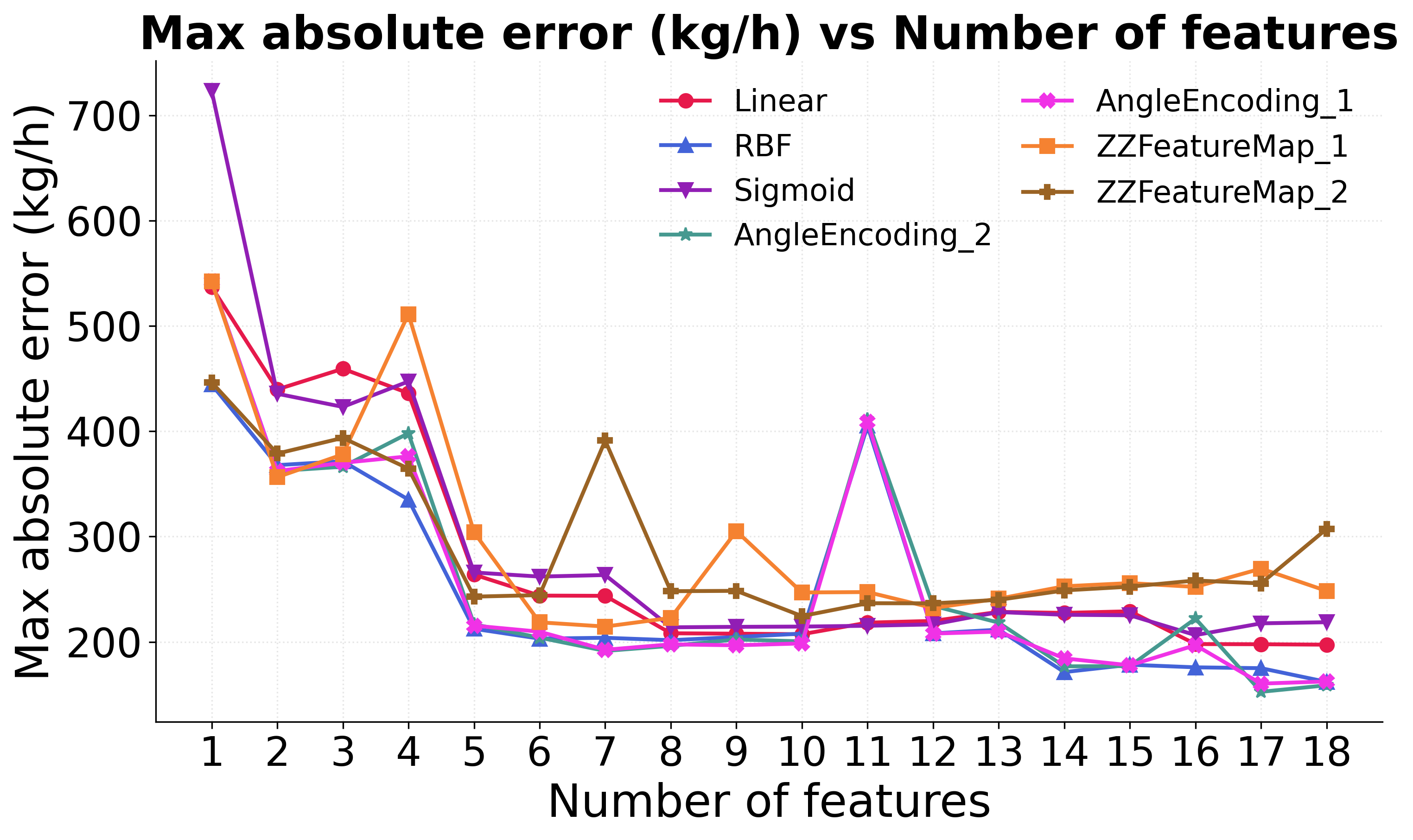}
\end{pfig}

\begin{figure}[!htbp]
\centering
\begin{minipage}[t]{0.48\textwidth}
\centering
\includegraphics[width=\linewidth]{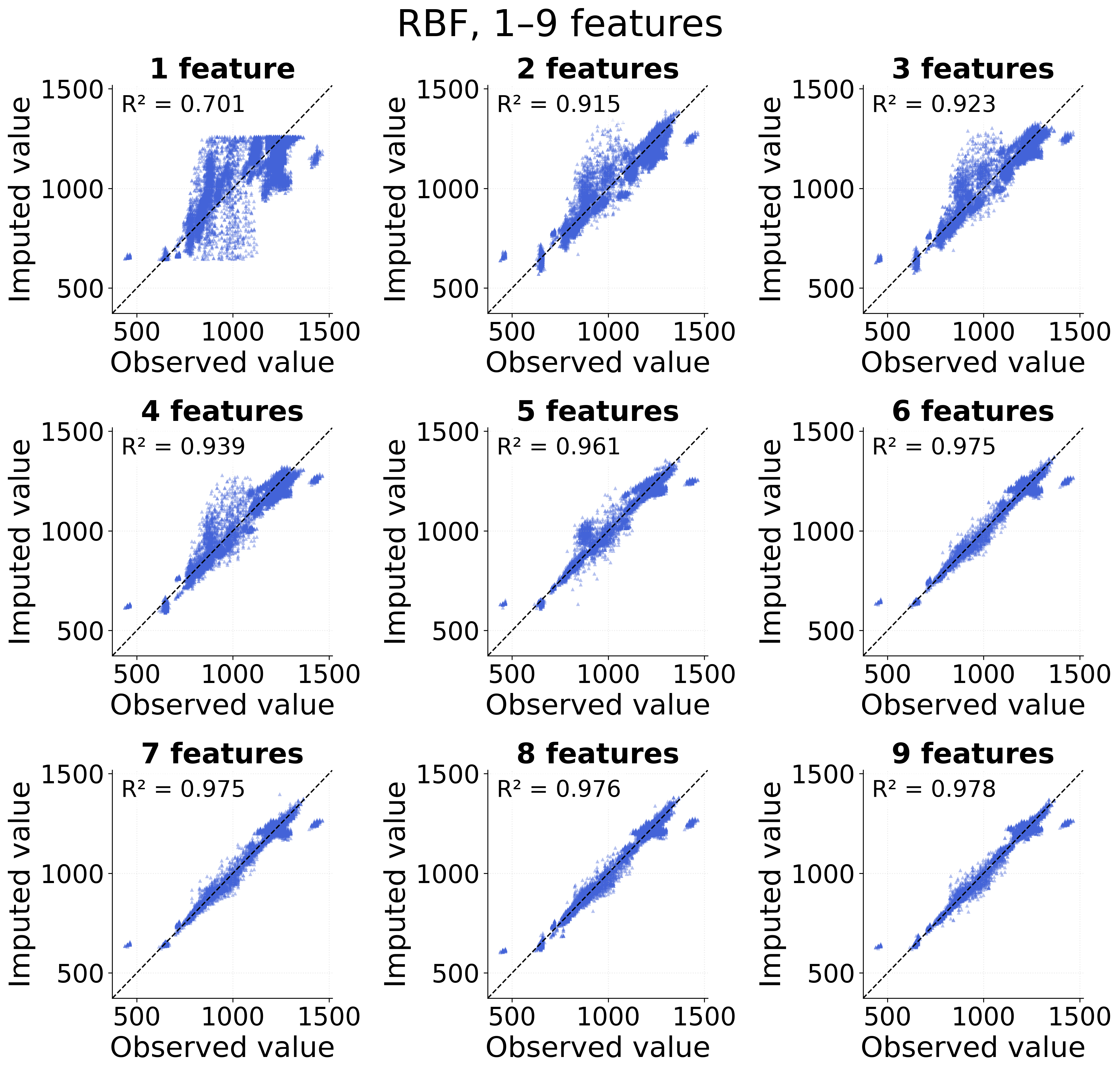}
\caption{Observed versus imputed fuel-flow values for the RBF kernel with one to nine input features, using 100\% of the dataset and 90\% missing target values. The dashed diagonal denotes perfect agreement.}
\label{fig:scalability-scatter-rbf-a}
\end{minipage}\hfill
\begin{minipage}[t]{0.48\textwidth}
\centering
\includegraphics[width=\linewidth]{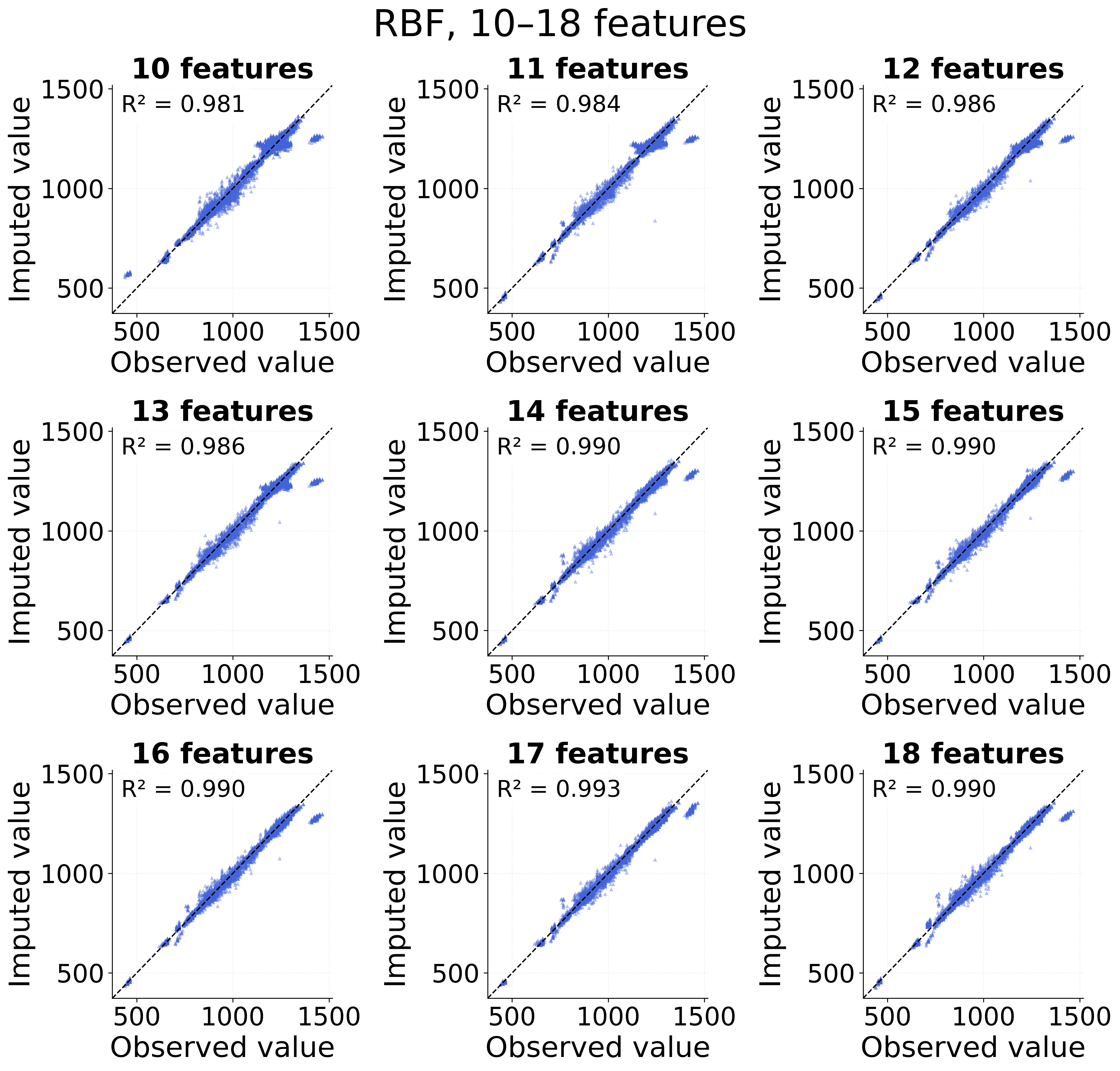}
\caption{Observed versus imputed fuel-flow values for the RBF kernel with 10 to 18 input features, using 100\% of the dataset and 90\% missing target values. The dashed diagonal denotes perfect agreement.}
\label{fig:scalability-scatter-rbf-b}
\end{minipage}
\end{figure}

Figures~\ref{fig:scalability-scatter-rbf-a} and~\ref{fig:scalability-scatter-rbf-b} show that the RBF predictions approach the identity line as features are added, with the largest gains occurring at lower dimensionalities. The $R^2$ values shown in the plots increase from 0.701 with one feature to 0.978 with nine. Across 10 to 18 features, they remain around 0.98--0.99, reaching 0.993 at 17 features. Thus, the 14-feature configuration already provides a close fit, while further gains are modest and not strictly monotonic.

\begin{figure}[!htbp]
\centering
\begin{minipage}[t]{0.48\textwidth}
\centering
\includegraphics[width=\linewidth]{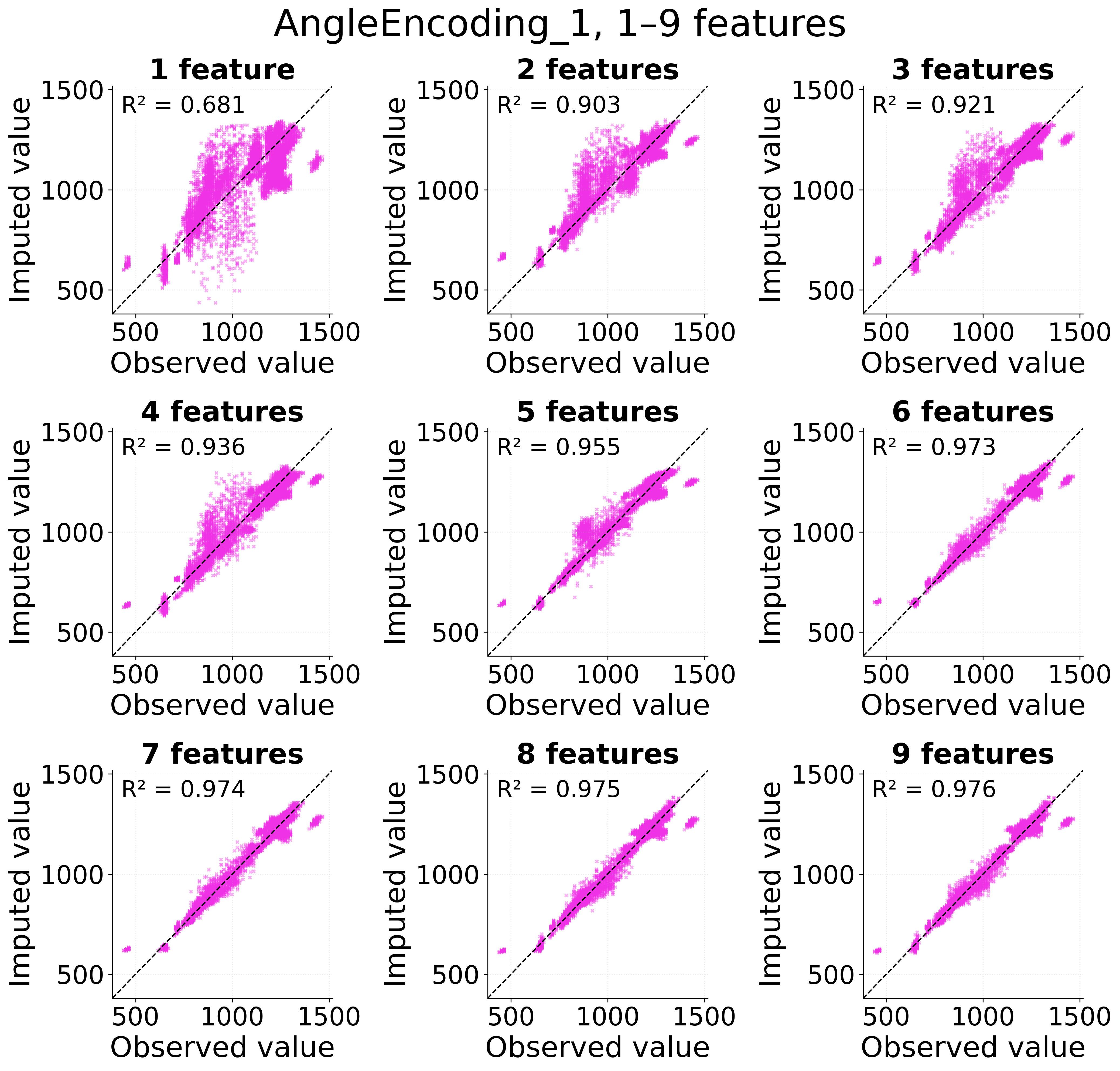}
\caption{Observed versus imputed fuel-flow values for AngleEncoding\_1 with one to nine features and matching qubit counts, using 100\% of the dataset and 90\% missing target values. The dashed diagonal denotes perfect agreement.}
\label{fig:scalability-scatter-ae1-a}
\end{minipage}\hfill
\begin{minipage}[t]{0.48\textwidth}
\centering
\includegraphics[width=\linewidth]{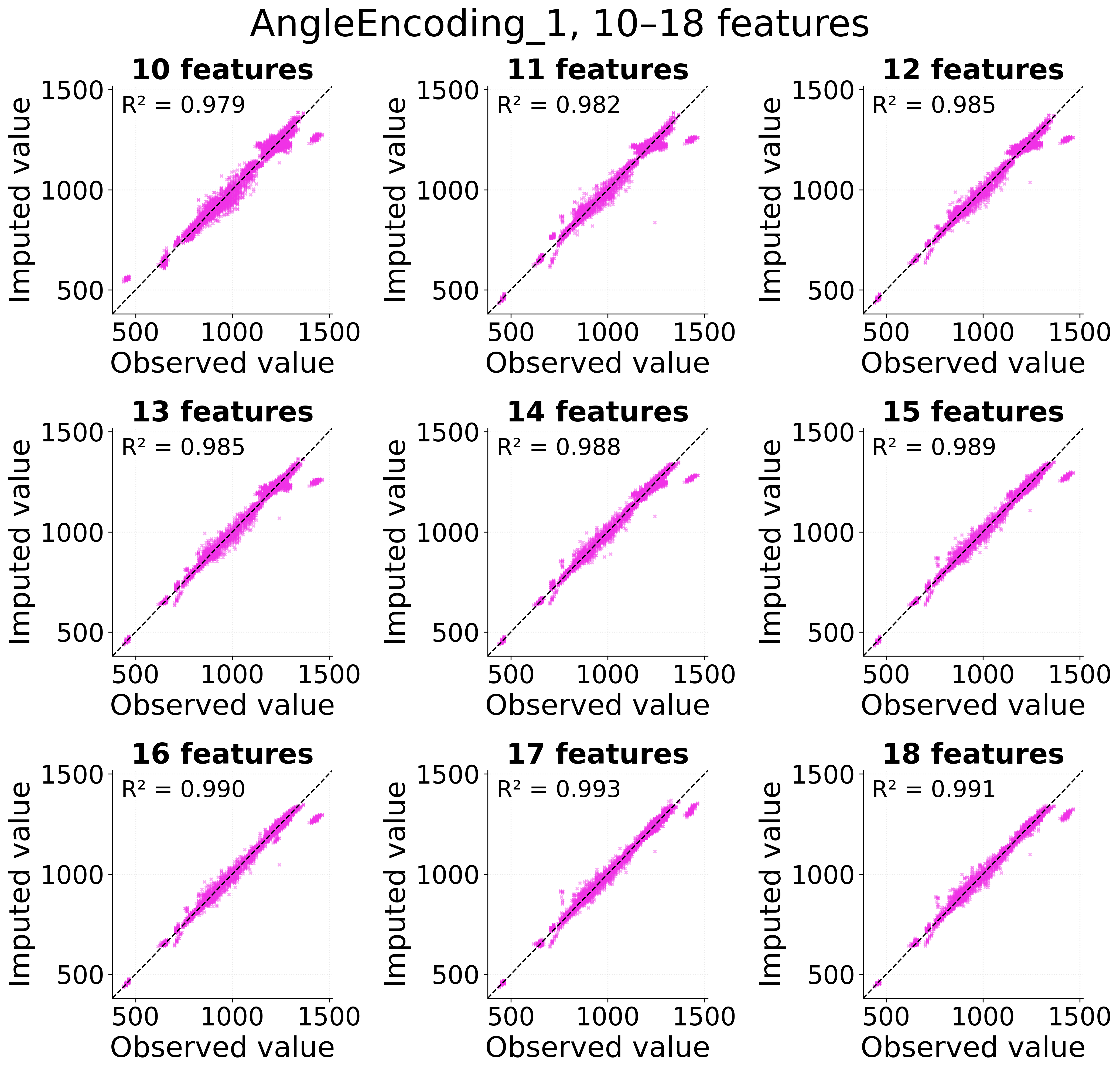}
\caption{Observed versus imputed fuel-flow values for AngleEncoding\_1 with 10 to 18 features and matching qubit counts, using 100\% of the dataset and 90\% missing target values. The dashed diagonal denotes perfect agreement.}
\label{fig:scalability-scatter-ae1-b}
\end{minipage}
\end{figure}

Figures~\ref{fig:scalability-scatter-ae1-a} and~\ref{fig:scalability-scatter-ae1-b} show a similar improvement for AngleEncoding\_1. Its $R^2$ increases from 0.681 with one feature to 0.976 with nine and reaches 0.988 at 14 features. Improvements continue at some higher dimensionalities, reaching 0.993 at 17 features, but remain modest compared with the gains observed when moving from the lowest feature counts. Considering this pattern alongside the statevector resource requirements supports using 14 features as a practical compromise.

Computational cost provides the other component of this decision. Increasing the qubit count enlarges the full statevector representation exponentially, so extending the complete experimental analysis to higher dimensionalities entails additional resource requirements. Given the limited and inconsistent predictive gains beyond 14 features, this dimensionality was retained as a practical compromise. The dimensionality plots establish the predictive component of this choice; measured runtimes for the selected configuration are compared in Section~\ref{subsec:results-representative}.

\subsection{Comparison under the selected representative configuration}
\label{subsec:results-representative}

The final stage combines the findings of the exploratory and scalability analyses. Unlike the scalability tests, which used the full dataset, this stage reduces the dataset to 60\% while retaining 90\% missing target values. Table~\ref{tab:method_comparison} reports this comparison with 14 input features, corresponding to 14 qubits for the quantum models. The 60\% subset was selected on the basis of the exploratory results, which showed sufficiently stable performance and low overfitting without requiring the full dataset. The 14-feature choice was guided by the full-dataset scalability analysis, in which further increases in dimensionality yielded limited and model-dependent gains. This combination retains a demanding missingness rate while avoiding a full-dataset comparison at higher dimensionalities. These choices define a representative, resource-conscious scenario, not a theoretical limit or a universally optimal configuration. Because the scalability and final stages use different dataset sizes, their scores should not be interpreted as repeated measurements under identical conditions. Runtime includes both training and inference.

\begin{ptable}[raw=true,span=double, pos=t]{Performance comparison using 60\% of the dataset, 90\% missing data, and 14 input features. Best values are highlighted in bold. Runtime includes training and inference; quantum runtimes correspond to statevector emulation, not quantum hardware.}{tab:method_comparison}{}
\centering

\small
\setlength{\tabcolsep}{3pt}
\begin{adjustbox}{max width=\textwidth}
\begin{tabular}{lccccc}
\hline
\textbf{Method} & \textbf{$R^2$ ($\pm$ std)} & \textbf{RMSE (kg h$^{-1}$) ($\pm$ std)} & \textbf{Runtime (s)} & \textbf{Relative overfitting} & \textbf{Max. error (kg h$^{-1}$)} \\
\hline
ZZFeatureMap\_1 & \bestvalue{0.9912 $\pm$ 0.0012} & \bestvalue{19.51 $\pm$ 1.38} & 1,439.76 & 0.00795 & 259.62 \\
rbf & 0.9896 $\pm$ 0.0010 & 21.18 $\pm$ 1.01 & 1.78 & 0.00288 & 235.43 \\
AngleEncoding\_1 & 0.9883 $\pm$ 0.0009 & 22.48 $\pm$ 0.81 & 1,197.67 & 0.00274 & 243.60 \\
ZZFeatureMap\_2 & 0.9882 $\pm$ 0.0011 & 22.63 $\pm$ 1.02 & 1,545.38 & 0.01134 & 256.21 \\
AngleEncoding\_2 & 0.9879 $\pm$ 0.0006 & 22.94 $\pm$ 0.52 & 1,175.73 & 0.00228 & 250.16 \\
poly & 0.9873 $\pm$ 0.0009 & 23.43 $\pm$ 0.82 & 88.46 & 0.00314 & 624.58 \\
linear & 0.9579 $\pm$ 0.0004 & 42.73 $\pm$ 0.25 & 1.77 & 0.00155 & 236.83 \\
sigmoid & 0.9555 $\pm$ 0.0007 & 43.97 $\pm$ 0.38 & \bestvalue{1.36} & \bestvalue{0.00055} & \bestvalue{226.65} \\
\hline
\end{tabular}
\end{adjustbox}
\end{ptable}

ZZFeatureMap\_1 achieves the strongest predictive performance, with the highest $R^2$ and lowest RMSE, but has a substantially longer runtime than the classical kernels. The RBF kernel follows closely in imputation performance while reducing runtime by approximately three orders of magnitude and maintaining a smaller relative overfitting value, making it the strongest overall accuracy--efficiency trade-off in this scenario. AngleEncoding\_1, AngleEncoding\_2, and ZZFeatureMap\_2 also achieve high imputation performance, although their runtimes remain above 1,100 seconds; among them, ZZFeatureMap\_2 exhibits the greatest relative overfitting. These higher quantum runtimes are an important practical consideration: in this setup, adding input features also adds qubits and enlarges the circuits, while quantum-kernel estimation requires repeated circuit evaluations. Consequently, the cost of executing many circuits can become substantial as circuit size grows. The polynomial kernel provides competitive average accuracy at a lower computational cost than the quantum models, but its markedly larger maximum error indicates occasional substantial prediction deviations. Finally, the linear and sigmoid kernels are the fastest and show the lowest relative overfitting values, but their lower $R^2$ and higher RMSE demonstrate that this efficiency is obtained at the expense of imputation performance.

\subsection{Summary of findings and limitations}
\label{subsec:results-summary}

The three experimental stages lead to the following findings:

\begin{itemize}
    \item The three-feature exploratory analysis establishes that both classical and quantum models can reconstruct missing fuel-flow values across a broad range of dataset sizes and MCAR missingness rates. Aggregate test $R^2$ values range from 0.918 to 0.945, and overfitting is generally low outside the most data-scarce conditions. These results characterise sensitivity to data availability rather than establish a definitive best model.
    \item The scalability analysis, conducted with 100\% of the dataset and 90\% missing target values, shows that increasing the feature set improves predictive performance substantially at low dimensionalities, but the gains become smaller and model-dependent around and beyond 14 features. This evaluates feature dimensionality at the highest missingness rate considered while retaining the full dataset. Some configurations improve further at 17 features, whereas others deteriorate. Fourteen features were therefore selected as a practical accuracy--resource compromise, not as a proven optimum or the maximum dimensionality tested.
    \item The final comparison combines the 14-feature choice from the full-dataset scalability analysis with the 60\% dataset subset supported by the exploratory analysis, retaining 90\% missing target values. The reduced dataset choice retains sufficiently low overfitting while avoiding the cost of using all available instances. This is a case-study-specific lower-data choice and should not be interpreted as a universal minimum data requirement.
    \item In the final comparison, ZZFeatureMap\_1 achieves the highest mean $R^2$ ($0.9912$) and lowest RMSE ($19.51$~kg~h$^{-1}$). The RBF kernel follows closely, with $R^2=0.9896$ and RMSE $21.18$~kg~h$^{-1}$, while requiring only $1.78$~s compared with $1{,}439.76$~s for ZZFeatureMap\_1. Thus, the strongest observed quantum accuracy comes with a substantial statevector-emulation runtime cost, and RBF provides a competitive accuracy--efficiency balance. These observations do not establish statistically significant superiority or computational quantum advantage.
    \item Low overfitting does not, by itself, imply the best imputation accuracy. In the final comparison, the linear and sigmoid kernels have the smallest relative overfitting values but larger RMSE than  the RBF, polynomial and quantum kernels. Likewise, the polynomial kernel has competitive average accuracy but the largest maximum error. Evaluating accuracy, generalisation, individual prediction deviations, and runtime together is therefore important.
\end{itemize}

Based on these findings, several limitations were also identified. These limitations motivate the following directions for future research:

\begin{itemize}
    \item The main scope of this study was to compare QSVR and SVR in a data-imputation regression task. To further assess the feasibility and potential benefits of QML, future studies should investigate additional state-of-the-art classical machine learning methods, as well as alternative QML approaches.
    \item The initial exploration used three features, the scalability analysis examined up to 18 features and qubits, and the final comparison used 14. These experiments were performed with exact statevector emulation, so the selected dimensionality and measured runtimes are implementation-dependent. Future work should evaluate broader feature sets, alternative ways of selecting features, and execution on quantum hardware, and report runtime and memory scaling across dimensionalities.
    \item The representative configuration was chosen from the exploratory and scalability results for this case study. Validation on additional vessels and operational regimes, under MAR and MNAR mechanisms as well as MCAR, is needed to assess whether the same data-availability and dimensionality choices remain suitable. Formal statistical testing is also required before claiming superiority of one approach over another.
    \item The present study focused on a regression-based data-imputation task. Future research should extend the analysis to other machine learning tasks, such as classification and clustering, as well as to other predictive scenarios, including nowcasting and forecasting.
    \item The QSVR models were compared directly with classical SVR models. Future work should also investigate hybrid quantum-classical architectures and assess whether combining quantum and classical components can provide improvements in predictive performance, computational efficiency, or scalability.

\end{itemize}

\section{Conclusions}
\label{sec:conclusions}

This work presented a comparative study of classical SVR kernels and QSVR models based on different quantum feature maps for the estimation of marine fuel consumption under varying levels of data availability and missing information. The evaluation considered multiple dataset sizes and missing-data proportions, allowing the robustness of both classical and quantum approaches to be assessed across a range of operating conditions representative of real-world maritime monitoring systems.

Across all evaluated scenarios, both classical and quantum approaches successfully captured the relationship between the input variables and the main-engine fuel-flow rate, maintaining strong predictive performance even under severe data scarcity conditions. Among the evaluated configurations, ZZFeatureMap\_1 achieved the highest average $R^2$ and the lowest aggregate imputation errors, while the RBF kernel consistently emerged as the strongest classical baseline with highly competitive results. In contrast, the linear and sigmoid kernels exhibited lower predictive capability, indicating that a sufficiently expressive kernel is required for this regression task. The scalability analysis further showed that most of the accuracy gain is obtained with the first few input features and that improvements beyond 14 features are limited and model-dependent. Generalisation remained stable across all methods, with only minor overfitting observed under the most challenging scenarios. In the three-feature exploratory analysis, computational costs depended on the model and data-availability scenario rather than showing a uniform difference between classical and quantum approaches. In the 14-feature comparison, however, quantum kernels required runtimes approximately 660 to 870 times longer than the RBF kernel, reflecting the exponential growth of emulation cost with the number of qubits. Median runtimes were used for the scenario-wise comparison because isolated timing peaks were attributed to transient computational-infrastructure overhead, making the mean less representative of typical execution time. These measurements reflect the CPU-based implementations and statevector emulation used in this study, not quantum-hardware performance or intrinsic limitations of QML methods.

Overall, the results indicate that quantum feature-map-based regression can achieve accuracy comparable to that of the best-performing classical kernels and, in certain scenarios, marginally outperform them. However, the observed performance differences remain relatively small and fall within the margins of experimental variability. Consequently, these findings should be interpreted as evidence of the viability and competitiveness of quantum kernel methods rather than evidence of statistically significant superiority. Furthermore, although the quantum models exhibit competitive predictive capabilities, the present results do not constitute evidence of quantum advantage in the computational sense, since all quantum kernels were evaluated through classical statevector emulation and, at 14 qubits, required substantially longer runtimes than the classical kernels.

Future work should extend the analysis by evaluating additional quantum feature maps, alternative QML models, and larger maritime datasets. Furthermore, the use of actual quantum hardware would provide a more representative assessment of computational performance and scalability. Finally, formal statistical significance testing and investigations into uncertainty quantification could provide further insight into the practical advantages and limitations of quantum-enhanced regression methods for maritime applications.

\textbf{Credit Authorship Contribution Statement }

\begin{enumerate}
    \item Daniel Maroto Sanchez: Conceptualization, Methodology, Software, Formal Analysis, Investigation, Writing - Original Draft, Writing - Review \& Editing, Visualization
    \item Christian Velasco-Gallego: Conceptualization, Methodology, Software, Formal Analysis, Investigation, Writing - Original Draft, Writing - Review \& Editing, Visualization
    \item Miriam Bastante Chichon: Conceptualization, Methodology, Software, Formal Analysis, Writing - Original Draft, Writing - Review \& Editing, Visualization
    \item Jaime Álvarez-Benayas: Methodology, Software, Formal Analysis, Investigation, Writing - Original Draft, Writing - Review \& Editing, Visualization
    \item Tsoulakos Nikolaos: Data Curation, Writing - Original Draft, Writing - Review \& Editing
    \item J. A. Bravo-Montes: Conceptualization, Investigation, Writing - Review \& Editing, Visualization

\end{enumerate}

\textbf{Declaration of generative AI use}
During the preparation of this manuscript, the authors used generative AI (Claude Sonnet 5 and GPT-6) for translation, grammar correction, and assistance in generating certain portions of the code (e.g., the plots in the results section). The authors have reviewed and edited the output and take full responsibility for the content of this publication.

\textbf{Conflicts of interest}
The authors declare that the authors Daniel Maroto Sanchez, Miriam Bastante Chichon, and J.A. Bravo-Montes are employed by Bull Spain and Tsoulakos Nikolaos is employed by Laskaridis Shipping Co. Ltd.

\bibliographystyle{plainnat}

\bibliography{references}

\appendix
\section{Additional plots for the exploratory analysis}
\label{app:exploratory-additional}

This appendix provides complementary exploratory results for the metrics whose overall trends are already represented in Section~\ref{subsec:results-exploratory}. The MAE and $R^2$ plots show patterns consistent with the RMSE results. The absolute generalisation-gap plot likewise supports the main-text discussion of relative overfitting, with no materially different interpretation.

Figure~\ref{fig:comparison-mae} shows that the performance differences are reflected in the typical absolute imputation error. Across most dataset sizes and missing-data fractions, ZZFeatureMap\_1 and the RBF kernel consistently achieve the lowest MAE values, whereas the linear and sigmoid kernels exhibit the highest errors. The remaining methods form an intermediate group with broadly comparable performance. The effect of missing-data proportion on MAE is generally limited, while dataset size has a more noticeable impact on performance variability. The largest uncertainty is observed for ZZFeatureMap\_2 in the smallest datasets, whereas variability decreases substantially as more training data become available. Overall, the relative ranking of the evaluated methods remains largely stable across the different data-availability scenarios.

\begin{pfig}[float=false,span=double]{Mean MAE on the imputed test set across dataset sizes and missing-data percentages. Shaded regions represent $\pm1$ standard deviation.}{fig:comparison-mae}
\centering
\includegraphics[width=0.9\textwidth,height=.74\textheight,keepaspectratio]{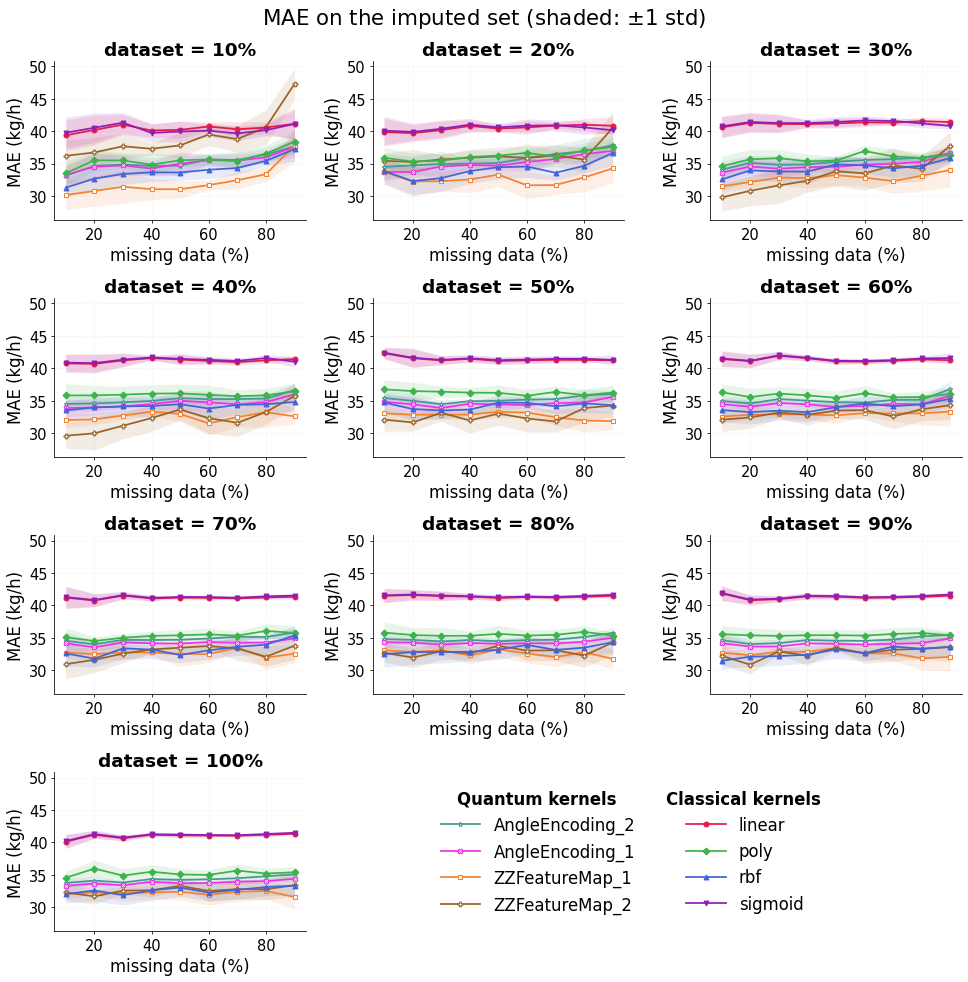}
\end{pfig}

Figure~\ref{fig:comparison-r2} shows that ZZFeatureMap\_1 consistently achieves the highest or near-highest test $R^2$ values across most dataset sizes and missing-data levels, closely followed by the RBF kernel. ZZFeatureMap\_2, poly, AngleEncoding\_1, and AngleEncoding\_2 form a second group with comparable performance, whereas the sigmoid kernel systematically yields the lowest $R^2$ values (approximately 0.92). Overall, the impact of increasing missing-data levels is relatively limited, particularly for medium and large datasets, where the curves remain nearly flat and the uncertainty bands are narrow. The largest performance degradation and variability are observed for ZZFeatureMap\_2 when only 10--20\% of the dataset is available, especially at high missing-data fractions.

\begin{pfig}[float=false,span=double]{Mean coefficient of determination on the imputed test set across dataset sizes and missing-data percentages. Shaded regions represent $\pm1$ standard deviation.}{fig:comparison-r2}
\centering
\includegraphics[width=.9\textwidth,height=.74\textheight,keepaspectratio]{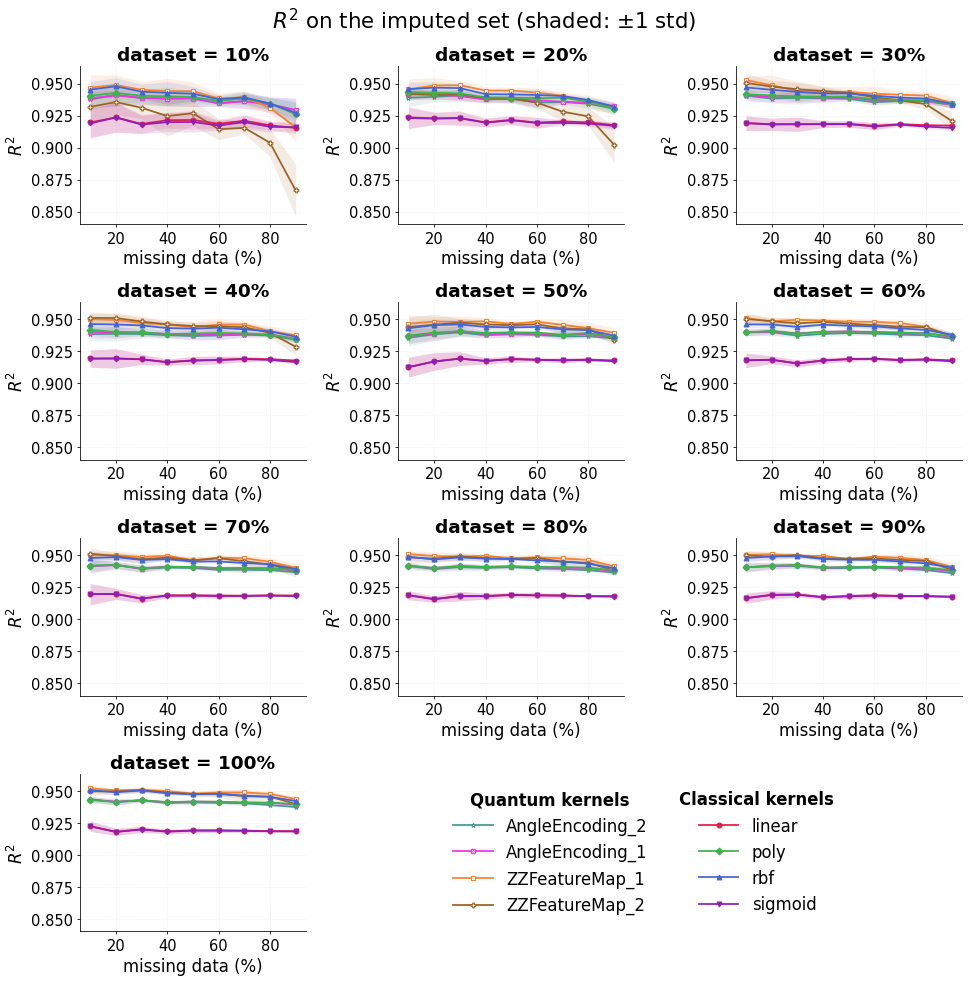}
\end{pfig}

Figure~\ref{fig:comparison-overfitting-gap} shows that the absolute generalisation gap remains close to zero for most methods and experimental conditions, indicating only minor differences between training and imputed-test performance. For dataset sizes above 30--40\%, the gap is generally stable and largely insensitive to the proportion of missing data, suggesting good generalisation across both classical and quantum approaches. A different behaviour emerges under severe data scarcity, where small datasets are combined with high levels of missing data. In these conditions, the generalisation gap increases noticeably for some methods, particularly ZZFeatureMap\_2, whose mean gap rises sharply as missingness exceeds 80\%. In contrast, the RBF kernel, ZZFeatureMap\_1, AngleEncoding\_1, and AngleEncoding\_2 maintain comparatively small gaps across most scenarios. Overall, the results indicate that overfitting remains limited for the majority of configurations, with substantial degradation appearing only in the most challenging data-availability conditions.

\begin{pfig}[float=false,span=double]{Absolute generalisation gap, $R^2_{\mathrm{train}}-R^2_{\mathrm{test}}$, across dataset sizes and missing-data percentages. Positive values indicate better performance on the observed training set.}{fig:comparison-overfitting-gap}
\centering
\includegraphics[width=0.9\textwidth,height=\textheight,keepaspectratio]{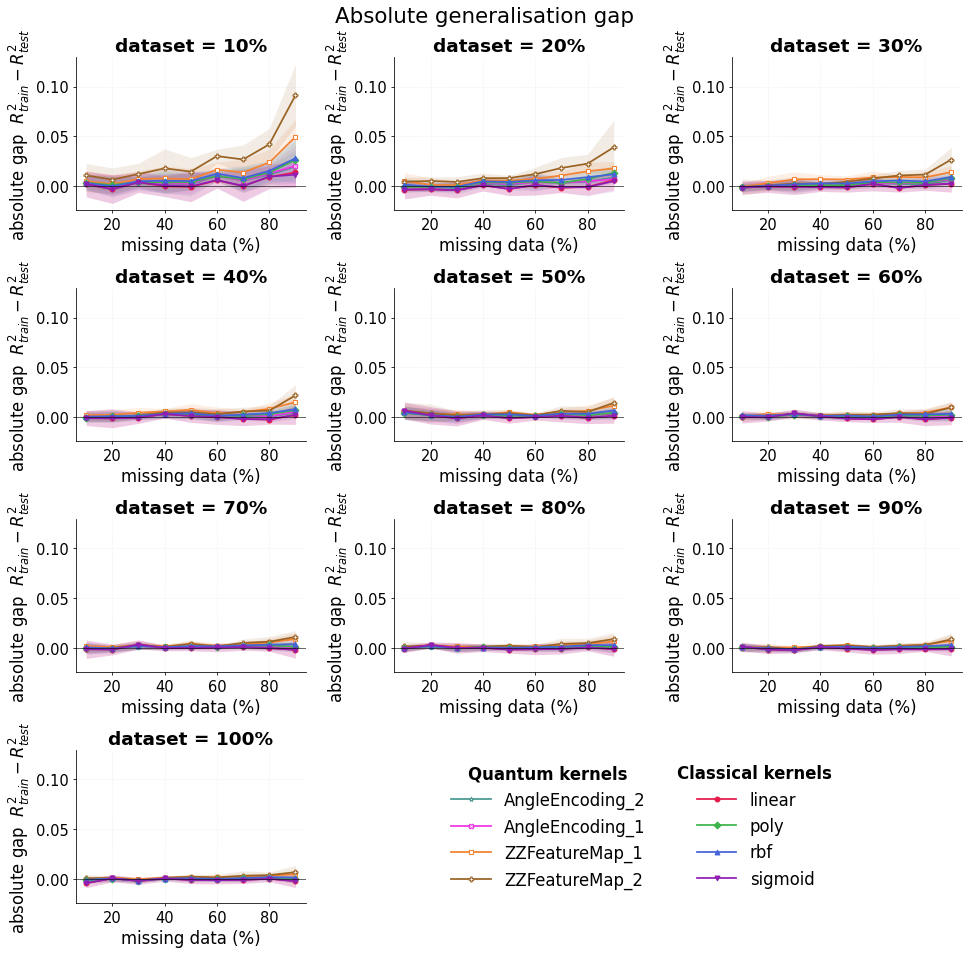}
\end{pfig}

\typeout{get arXiv to do 4 passes: Label(s) may have changed. Rerun}

\end{document}